\documentclass[preprint,aps,nofootinbib]{revtex4}

\usepackage{dcolumn}
\usepackage{bm}
\usepackage{epsfig}
\usepackage{graphicx,subfigure}

\usepackage[bookmarks=true]{hyperref}

\usepackage{amsmath}
\usepackage{amsfonts}
\usepackage{amssymb}
\usepackage{slashed}

\usepackage{caption}

\usepackage{cprotect}

\usepackage{amsmath}
\usepackage{amsfonts}
\usepackage{amssymb}
\usepackage{color}

\def\be{\begin{equation}}
\def\ee{\end{equation}}
\def\bea{\begin{eqnarray}}
\def\eea{\end{eqnarray}}

\def\ss2l{SS2$\ell$}
\def\3l{3$\ell$}

\def\slashchar#1{\setbox0=\hbox{$#1$}           
   \dimen0=\wd0                                 
   \setbox1=\hbox{/} \dimen1=\wd1               
   \ifdim\dimen0>\dimen1                        
      \rlap{\hbox to \dimen0{\hfil/\hfil}}      
      #1                                        
   \else                                        
      \rlap{\hbox to \dimen1{\hfil$#1$\hfil}}   
      /                                         
   \fi}

\begin{document}
\title{Effective Field Theory for Higgs  Plus Jet Production}
\vspace*{1cm}

\author{\vspace{1cm} S.~Dawson$^{\, a}$, I.~M.~Lewis$^{\, a}$ and   Mao Zeng$^{\, b}$ }

\affiliation{
\vspace*{.5cm}
  \mbox{$^a$Department of Physics,\\
  Brookhaven National Laboratory, Upton, N.Y., 11973,  U.S.A.}\\
 \mbox{$^b$ C.N. Yang Institute for Theoretical Physics\\
 Stony Brook University,}
 \mbox{
 Stony Brook, N.Y., 11794, U.S.A.}\\
\vspace*{1cm}}

\begin{abstract}
\vspace*{0.5cm}
We use an effective field theory (EFT) which includes all possible gluon-Higgs dimension-5 and dimension-7 operators to
study Higgs boson plus jet production in next-to-leading order QCD. The EFT sheds light on the effect of a finite top quark mass as well as any Beyond-the-Standard Model (BSM) modifications of Higgs-gluon effective couplings.  In the gluon channel, the accuracy of the heavy-top approximation for differential distributions arises from the non-interference between the helicity amplitudes of the $G^3 h$ and $G^2 h$ operators in the $m_h < p_T$ limit at lowest order.  One dimension-7 operator involving quark bilinears, however, contributes significantly at high $p_T$, and potentially offers a channel for seeing BSM effects.  One-loop renormalization of these operators is determined, allowing resummation of large logarithms via renormalization group running. NLO numerical results at the LHC are presented, which include $\mathcal O ( 1/m_t^2 )$ contributions in the SM limit.
\end{abstract}

\maketitle
\tableofcontents
\section{Introduction}

The recently discovered Higgs boson has all the generic characteristics of a  Standard Model
 Higgs boson and
measurements of the production and decay rates
agree to the $10-20\%$ level with Standard Model  (SM) predictions \cite{ATLAS-CONF-2014-009,CMS-PAS-HIG-14-009,Dittmaier:2011ti,Dittmaier:2012vm}.  The largest contribution to Standard Model
Higgs boson production comes from gluon fusion through a top quark loop
and testing the nature of this Higgs-gluon
interaction probes the mechanism of electroweak symmetry breaking at high scales.
In models with new physics, the gluon fusion rate can be altered by new particles interacting
in the loop which contribute to an effective dimension-5 operator \cite{Kniehl:1995tn,Spira:1995rr,Dawson:1990zj},
\begin{equation}
\mathcal L_5={\hat C}_1 G^{\mu\nu,A}G_{\mu\nu}^A h\, .
\label{leff5}
\end{equation}

For example,   in composite models ${\hat C}_1$ is 
changed from its SM value by small contributions of ${\cal O}(v^2/f^2)$, where $f$ is a ${\rm TeV}$ scale parameter corresponding
to the composite scale \cite{ArkaniHamed:2002qy,Contino:2003ve,Low:2010mr}.  
Similarly, supersymmetric models alter the $ggh$ coupling due to the contributions
of new particles such as squarks in the loops and  also
by changes in the Higgs-fermion couplings \cite{Carena:2013qia,Carena:2013iba,Dittmaier:2011ti,Dittmaier:2012vm}.  
The measurement of gluon fusion by itself can only measure a combination 
of ${\hat C}_1$ and the top quark Yukawa coupling, but cannot distinguish between the two
potential new physics effects \cite{Grojean:2013nya,Azatov:2013xha,Low:2009di}. 

The high $p_T$ production of the Higgs boson through the process 
$pp\rightarrow h$+jet is particularly sensitive to new contributions to the 
Higgs  gluon effective coupling \cite{Grojean:2013nya,Buschmann:2014twa,Azatov:2013xha,Schlaffer:2014osa}.  
This is straightforward to demonstrate in top partner models where at low energy there is a cancellation
between the SM top and the top partner contributions to the gluon fusion rate for Higgs production, making
it extremely difficult to observe top partner physics in this channel \cite{Low:2009di,Azatov:2011qy,Dawson:2012di}.  The effects of top partners become
apparent, however, when kinematic distributions for 2-particle final states, such as 
double Higgs production \cite{Gillioz:2012se,Dawson:2012mk}, or
Higgs plus jet production \cite{Banfi:2013yoa}, are analyzed.   
The measurement of Higgs plus jet production offers the possibility to untangle new physics effects contributing
to the Higgs-gluon effective interactions from beyond the SM
(BSM) contributions to the Higgs-fermion Yukawa couplings.

The strong Higgs-gluon-light quark 
interactions can be parameterized through $SU(3)$ invariant effective dimension-$5$
and dimension-$7$ operators coupling the Higgs boson to partons, which are well known \cite{Neill:2009tn,Harlander:2013oja}.
The dimension -$5$ operator of Eq. \ref{leff5}
has been used to calculate SM Higgs production through NNLO \cite{Harlander:2002wh,Ravindran:2003um,Anastasiou:2002yz},
along with the Higgs $p_T$ distribution \cite{Anastasiou:2005qj,Catani:2007vq,Ravindran:2002dc}. 
At NLO, the total rate can be compared with an analytic result with exact top and bottom quark
 mass dependence \cite{Spira:1995rr},
while at NNLO, the effective theory calculation has been compared numerically with the calculation in the full 
theory \cite{Harlander:2009mq,Pak:2009dg}. In
both instances, the dimension-5 operator gives an extremely accurate approximation to the total rate for Higgs production
through gluon fusion.    The Lagrangian of Eq. \eqref{leff5} corresponds to the $m_t\rightarrow\infty$ limit of the SM, and ${\hat C}_1$
has been determined to ${\cal O }(\alpha_s^3)$ in the SM \cite{Chetyrkin:1997un,Schroder:2005hy,Chetyrkin:2005ia,Kramer:1996iq}.

 In this paper, we examine the effect of both the dimension-$5$ and dimension -$7$ gluon-Higgs 
 operators on Higgs  plus jet production at NLO QCD .  We present analytic formulas which can be applied 
to arbitrary models of new physics.  The effects of these operators on the Higgs $p_T$ 
distribution has been studied numerically at lowest order in 
Ref. \cite{Harlander:2013oja}.  
The Standard Model rate for Higgs +jet is known analytically at order ${\cal O}(\alpha_s^3)$ \cite{Ellis:1987xu,Baur:1989cm}, 
 while the NLO rate is known analytically in the $m_t\rightarrow\infty$ 
limit, \cite{deFlorian:1999zd,Glosser:2002gm,Ravindran:2002dc} which 
corresponds to the contribution from ${\hat C}_1$.   
Finite top mass effects in SM NLO corrections have been obtained as a
numerical expansion in $1/m_t^2$ \cite{Harlander:2012hf,Grazzini:2013mca,Bagnaschi:2011tu, Neumann:2014nha}, and agree with the $m_t\rightarrow\infty$
limit only for small Higgs transverse momentum, $p_T \le 150~$GeV.  
The electroweak contributions are studied in \cite{Keung:2009bs}. The NNLO total cross section in the $m_t\rightarrow\infty$ limit for the $gg$ channel is known \cite{Boughezal:2013uia}  while the corresponding results for other partonic channels have been obtained in the threshold approximation \cite{Becher:2014tsa, Becher:2013vva, Huang:2014mca}.
For Higgs production in association with more than one jet, exact $m_t$ dependence is known for two and three jets at leading order \cite{DelDuca:2001eu, DelDuca:2001fn, Campanario:2013mga}, while $m_t \to \infty$ results are available at NLO for two and three jets \cite{vanDeurzen:2013rv, Cullen:2013saa}.

In Section \ref{efflag}, we discuss the effective Higgs-gluon  effective
Lagrangian, and in Section \ref{lores} we review the lowest order results
for Higgs plus jet production
in the dimension-7 effective field theory  (EFT). The renormalization of the dimension-7 effective Lagrangian coefficients is discussed in Section \ref{rencond}. 
  Sections \ref{lhcres} and \ref{nloreal} contain analytic results for Higgs plus jet production at NLO using the dimension-5
and dimension-7 contributions to the EFT, with the real emission corrections presented as heclity amplitudes using the conventions in \cite{Dixon:1996wi, Peskin:2011in}. The behavior of tree amplitudes in the massless Higgs limit, $m_h^2 < (p_T^2, s, -t, -u)$, is discussed.   As a by-product of our calculation, we obtain the ${\cal O} (1/m_t^2)$ contributions to the SM rate, modulo the non-logarithmic terms in the NLO matching coefficients in Eqs. \eqref{c3hat},\eqref{c5hat} which will be derived in a forthcoming work.
   Numerical results for the LHC are presented in Section \ref{lhcpheno}, and some
conclusions given in Section \ref{concs}.

\section{Effective Lagrangian}
\label{efflag}
\subsection{Higgs-gluon-quark interaction}
The calculations of Higgs production from gluon fusion are greatly simplified by using an effective Lagrangian
where heavy particles, such as the top quark, are integrated out.  
The $SU(3)$ invariant effective Lagrangian which parameterizes the CP-conserving Higgs -gluon- light quark
strong
interactions is,
\begin{equation}
\mathcal L_{\rm eff}= \hat C_1 O_1+{1\over \Lambda^2}\Sigma_{i=2,3,4,5} \hat C_i O_i+{\cal O}\biggl({1\over \Lambda^4}\biggr)
\, .\label{ldef}
\end{equation}
For SM Higgs production, $\Lambda = m_t$ is either the $\overline{\rm MS}$ running mass or the pole mass, depending on whether the $\overline{\rm MS}$ scheme or the pole scheme is used to calculate the matching coefficients, ${\hat C}_i$. For BSM scenarios, $\Lambda$ is the scale at which BSM physics generates 
 contributions to ${\hat C}_i$.

At dimension-5, the unique operator is
\begin{equation}
O_1=G_{\mu\nu}^A G^{\mu\nu,A}h\, ,
\end{equation}
where $G^A_{\mu\nu}$ is the gluon field strength tensor.  We consider only models with
a single scalar Higgs boson, although our results can be trivially generalized to the case with 
multiple scalars. 
In the SM, the coefficient, $\hat C_1$, is, to ${\cal O}\left( \alpha_s^2 \right)$ \cite{Dawson:1990zj,Spira:1995rr},
\begin{eqnarray}
\hat C_1(\mu_R)^{\rm SM,\,\overline{MS}}&=&{\alpha_s(\mu_R)\over 12\pi v}
\biggl\{
1 +{\alpha_s(\mu_R)\over 4 \pi}
\biggl[5C_A-3C_F\biggr]\biggr\}\, ,\label{c1hat}
\end{eqnarray}
where $C_A=N_c=3$ , $C_F={N_c^2-1\over 2N_c}={4\over 3}$, $v=246~{\rm GeV}$, and $\mu_R$ is an arbitrary
renormalization scale of ${\cal O}(m_h)$.

The dimension-7 operators, needed for gluon fusion production of Higgs, are \cite{Neill:2009tn,Harlander:2013oja,Buchmuller:1985jz},
\begin{eqnarray}
O_2&=& D_\sigma G^A_{\mu\nu}D^\sigma G^{A,\mu\nu} \,h \label{op2}
\\
O_3&=& f_{ABC}G_\nu^{A,\mu} G_\sigma^{B,\nu}G_\mu^{C,\sigma} \,h \label{op3}
\\
O_4&=&g_s^2\, \Sigma_{i,j=1}^{n_{lf}} {\overline \psi}_i\gamma_\mu T^A \psi_i \,
 {\overline \psi}_j\gamma^\mu T^A \psi_j \,h \label{op4}
\\
O_5&=& g_s \Sigma_{i=1}^{n_{lf}} 
G_{\mu\nu}^A D^\mu\, {\overline \psi}_i\gamma^\nu T^A\psi_i \,h \, ,
\label{op5}
\end{eqnarray}
where our convention for the covariant derivative is $D^\sigma=\partial^\sigma -ig_s T^A G^{A,\sigma}$,
$Tr(T^AT^B)={1\over 2}\delta_{AB}$
and $n_{lf}=5$ is the number of light fermions.
 The operators $O_1$, $O_2$ and $O_3$ are the only ones that are needed in pure QCD
 ($n_{lf}=0$). 
 In the presence of light quarks, we also need $O_4$ and $O_5$ which are related by the equations of motion
 (eom)  to gluon-Higgs operators\footnote{In our study, only gluons directly interact with the Higgs via a top quark loop or some BSM heavy particle, while quark-Higgs coupling is mediated by gluons.}
 \begin{eqnarray}
 O_4\mid_{eom}&\rightarrow &  D^\sigma G^A_{\sigma\nu}D_\rho G^{A,\rho\nu} h \equiv O_4'
 \nonumber \label{op4prime} \\
 O_5\mid_{eom} &\rightarrow & G^A_{\sigma\nu}D^\nu D^\rho G_\rho^{A,\sigma} h \equiv O_5' \, .
\label{op5prime}
\end{eqnarray} 
Since $O_4$ involves 4 light fermions, the operator contributes to Higgs plus jet production only starting at NLO, in
the  real-emission processes involving two incoming fermions and two outgoing fermions.

The SM coefficient, $\hat C_2^{\rm SM}$, can be found from the leading ${1\over m_t^2}$ terms in 
the NLO calculation of $gg\rightarrow h$ \cite{Dawson:1993qf}, in the ${\overline {MS}}$ scheme,
\begin{equation}
\hat C_2^{\rm{SM},\,\overline{\rm MS}}(\mu_R)= -{7\alpha_s(\mu_R)\over 720\pi v}\biggl\{ 1+
{\alpha_s(\mu_R)\over \pi}\biggl[ 
{29\over  84}C_A+{19\over 21}C_F+{3\over 2}
C_F\ln\biggl({m_t^2\over\mu_R^2}\biggr)\biggr]\biggr\}. \label{c2hat}
\end{equation}

For the remaining SM coefficients, we present only the LO contributions
along with the ${\alpha_s}\ln(m_t^2/\mu_R^2)$ contributions which can be deduced from the renormalization group equations in Section \ref{rencond}.
\footnote{The SM matching coefficients are given in Ref. \cite{Neill:2009tn}, but we found discrepancies at NLO. The $C_A \ln (m_t^2 / \mu_R^2)$ terms in our results are one half the values in \cite{Neill:2009tn}. Our results are consistent with the $O_3$ anomalous dimension found in \cite{Gracey:2002he} and the $O_5$ anomalous dimension we calculate in Section \ref{rencond}.  The non-logarithmic terms in the NLO matching coefficients, ${\hat C}^{(1)}_3$ and ${\hat C}^{(1)}_5$, will be discussed in a forthcoming work. In this study we will set 
${\hat C}^{(1)}_3$ and ${\hat C}^{(1)}_5$ to zero. Also, in Ref. \cite{Neill:2009tn} the matching is done off-shell, so the operator equivalence relation of Eq. \eqref{op5prime} cannot be used. As a result, in our convention the NLO value for $\hat C_5$ is different. The LO coefficients are in agreement with Refs.
\cite{Neill:2009tn,Harlander:2013oja}, once the differing sign conventions are accounted for.}
\begin{align}
\hat C_3^{\rm{SM},\,\overline{\rm MS}}(\mu_R)&= {g_s(\mu_R) 
\alpha_s(\mu_R)\over 60\pi v}\biggl\{ 1+{\alpha_s(\mu_R)\over \pi}
\left[ \hat C_3^{(1)} + \left( {1\over 4}C_A +{3\over 2}C_F \right) \right.
\left. \ln\biggl({m_t^2\over \mu_R^2}\biggr) \right] \biggr\} \label{c3hat}
\\
\hat C_4^{\rm{SM},\,\overline{\rm MS}}(\mu_R)&={\alpha_s(\mu_R)\over 360\pi v} + \mathcal O \left( \alpha_s^2 (\mu_R) \right) \label{c4hat}
\\
\hat C_5^{\rm{SM},\,\overline{\rm MS}}(\mu_R)&= {\alpha_s(\mu_R)\over 20\pi v}\biggl\{ 1+{\alpha_s(\mu_R)\over \pi}
\left[\hat C_5^{(1)} + \left( -{121\over 216}C_A + {59\over 54} C_F \right)
\ln\biggl({m_t^2\over \mu_R^2}\biggr) \right] \biggr\}\, . \label{c5hat}
\end{align}
Because the $O_4$ contribution starts at NLO for Higgs plus jet production, we have only presented the LO value for $\hat C_4$.  Since the above matching coefficients are presented in the $\overline{\rm MS}$ scheme, the top mass $m_t$ in Eq. \eqref{c3hat}-\eqref{c5hat}, as well as in Eq. \eqref{ldef}, should be taken as the $\overline{\rm MS}$ running top mass evaluated at the renormalization scale $\mu_R$.

To use the $\mu_R$-independent constant parameter $ 1/(m_t^{\rm pole})^2 $ as the EFT power expansion parameter in Eq. \eqref{ldef}, in line with the usual language for EFTs, we substitute into Eq. \eqref{ldef} the 
relation \cite{Melnikov:2000qh},
\begin{align}
m_t^{\overline{\rm MS}} (\mu_R) &= m_t^{\rm pole} \left\{ 1 - \frac{C_F \alpha_s (\mu_R)}{\pi} \left[1 - \frac 3 4 \ln \left( \frac{m_t^2}{\mu_R^2} \right) \right] + \mathcal O(\alpha_s^2) \right\},
\end{align}
which gives,
\begin{align}
\hat C_1^{\rm SM, pole}(\mu_R) &= \hat C_1^{{\rm SM},\overline{\rm MS}}(\mu_R), \label{c1hatpole} \\
\hat C_2^{\rm SM, pole}(\mu_R)&= -{7\alpha_s(\mu_R)\over 720\pi v}\biggl\{ 1+
{\alpha_s(\mu_R)\over \pi}\biggl[ 
{29\over  84}C_A+{61\over 21}C_F \biggr]\biggr\}, \label{c2hatpole} \\
\hat C_3^{\rm SM, pole}(\mu_R)&= {g_s(\mu_R)\alpha_s(\mu_R)\over 60\pi v}\biggl\{ 1+{\alpha_s(\mu_R)\over \pi}
\left[ \hat C_3^{(1)} + 2 C_F + {1\over 4}C_A   \right.
\left. \ln\biggl({m_t^2\over \mu_R^2}\biggr) \right] \biggr\} \label{c3hatpole}
\\
\hat C_4^{\rm SM, pole}(\mu_R)&={\alpha_s(\mu_R)\over 360\pi v} + \mathcal O \left( \alpha_s^2 (\mu_R) \right) \label{c4hatpole}
\\
\hat C_5^{\rm SM, pole}(\mu_R)&= {\alpha_s(\mu_R)\over 20\pi v}\biggl\{ 1+{\alpha_s(\mu_R)\over \pi}
\left[{\hat C}_5^{(1)} + 2 C_F + \left( -{121\over 216}C_A - {11\over 27} C_F \right)
\ln\biggl({m_t^2\over \mu_R^2}\biggr) \right] \biggr\}\, . \label{c5hatpole}
\end{align}

The Feynman rules corresponding to Eq. \ref{ldef} 
can be found in a straightforward manner. For most  of our calculations, we will use the pure-gluon operators $O_4'$ and $O_5'$ in Eq. \eqref{op4prime}  instead of $O_4$ and $O_5$ in Eqs. \eqref{op4} and \eqref{op5}, so that the Feynman diagrams for 
Higgs plus jet production from the dimension-7 operators are identical to those from the dimension-5 operator $O_1$. The $O_3$ vertices involve at least 3 gluons, while 2 gluons suffice for the other operators.

There are $2$ possible tensor structures \cite{Pasechnik:2006du} for the 
off-shell $g^{A,\mu}(p_1) g^{B,\nu}(p_2)h(p_3)$ vertex,
\begin{eqnarray}
T_1^{\mu\nu}&\equiv & g^{\mu\nu}p_1\cdot p_2-p_1^\nu p_2^\mu\nonumber \\
T_2^{\mu\nu}&\equiv &
p_1^\mu p_2^\nu
- p_2^\mu p_2^\nu {p_1^2\over p_1\cdot p_2}
-p_1^\mu p_1^\nu{p_2^2\over p_1\cdot p_2} +
p_1^\nu p_2^\mu{p_1^2p_2^2\over (p_1\cdot p_2)^2}\, .
\end{eqnarray}
The Lagrangian of Eq. \ref{ldef} has the off-shell Feynman rule,
\begin{eqnarray}
ggh:&&-i\delta_{AB} 
\biggl[T_1^{\mu\nu}X_1(p_1,p_2)+T_2^{\mu\nu}X_2(p_1,p_2)\biggr]\nonumber \\
X_1(p_1,p_2)&=& \biggl\{ 4\hat C_1-{\hat C_2\over \Lambda^2}4p_1\cdot p_2
-{\hat C_4\over \Lambda ^2}\biggl({2p_1^2p_2^2\over p_1\cdot p_2}\biggr)
+{\hat C_5\over \Lambda^2}(p_1^2+p_2^2)\biggr\}\nonumber \\
X_2(p_1,p_2)&=& -2p_1\cdot p_2{\hat C_4\over \Lambda^2}\, .
\end{eqnarray} 
The Feynman rules for the off-shell  $g(p_1^{A,\mu})g(p_2^{\nu,B}) g(p_3^{\rho,C})h(p_4)$ vertex
(with all momenta outgoing) are,\footnote{We omit the $\hat C_4$ $gggh$ vertex because this vertex does not contribute
to Higgs +jet at NLO.}
\begin{eqnarray}
O_1:&&-4\hat C_1 g_s f_{ABC}\biggl\{ -g^{\mu\nu} (p_1-p_2)^\rho+g^{\mu\rho}
(p_1-p_3)^\nu+g^{\nu\rho}(p_3-p_2)^\mu\biggl\}
\nonumber \\
O_2:&&-
4{\hat C_2\over \Lambda^2} g_s f_{ABC}
\biggl\{
  {\cal A}^{\mu\nu\rho}(p_1,p_2,p_3)
+{\cal A}^{\nu\rho\mu}(p_2,p_3,p_1)
+{\cal A}^{\rho\mu\nu}(p_3,p_1,p_2)\biggr\}
\nonumber \\
O_3:&&-6{\hat C_3\over \Lambda ^2}f_{ABC}{ Y}_0^{\mu\nu\rho}(p_1,p_2,p_3)
\nonumber \\
O_5:&&-g_s{\hat C_5\over \Lambda^2}\biggl\{f_{ABC}\biggl[
-g^{\mu\nu}p_1^\rho\biggl(
p_1^2+p_2^2+p_3^2-2p_1\cdot p_2-4p_2\cdot p_3\biggr)
\nonumber \\ &&
+2p_1^\nu p_2^\rho p_3^\mu+p_1^\nu p_1^\rho p_3^\mu-p_2^\mu p_2^\rho p_3^\nu\biggr]
+5{\hbox{~permutations}}\biggr\} \, ,
\end{eqnarray}
where 
\begin{eqnarray}
{Y}_{0}^{\mu\nu\rho}(p_1,p_2,p_3) & =&\left(p_{1}^{\nu}g^{\rho\mu}-p_{1}^{\rho}g^{\mu\nu}\right)p_2\cdot p_3
+\left(p_{2}^{\rho}g^{\mu\nu}-p_{2}^{\mu}g^{\nu\rho}\right)p_1\cdot p_3\nonumber \\
 & &\quad+\left(p_{3}^{\mu}g^{\nu\rho}-p_{3}^{\nu}g^{\rho\mu}\right)p_1\cdot p_2+p_2^\mu
 p_{3}^{\nu}p_{1}^{\rho}-p_{3}^{\mu}p_{1}^{\nu}p_{2}^{\rho}\nonumber \\
{\cal A}^{\mu\nu\rho}(p_1,p_2,p_3)&=&
(p_1-p_2)^\rho
T_1^{\mu\nu}(p_1,p_2)+p_1\cdot p_2
\biggl[ X_0^{\mu\nu\rho}(p_1)-X_0^{\nu\mu\rho}(p_2)\biggr]
\nonumber \\
X_0^{\mu\nu\rho}(p)&=&g^{\mu\nu}p^\rho-g^{\mu\rho}p^\nu\, .
\end{eqnarray}.

\subsection{Alternative operator basis}
\label{subsec:alternativeBasis}
In the previous section, we used the basis  of Eqs. \eqref{op2}-\eqref{op5} 
to describe the dimension-7 operators. Here we define another dimension- 7 operator,
\begin{equation}
O_6 = -D^{\rho} D_{\rho} \left( G_{\mu\nu}^A G^{\mu\nu,A} \right) h = m_h^2 O_1,
\label{opRelation}
\end{equation}
where the last equal sign is only valid for on-shell Higgs production, which will be assumed for the rest of this section. Using the Jacobi identities, without using the  equations of motion, we have the operator identity
\begin{equation}
O_6 = m_h^2 O_1 = -2 O_2 + 4 g_s O_3 + 4 O_5.
\label{oprel}
\end{equation}
Therefore, we can choose $O_6 = m_h^2 O_1$, $O_3$, $O_4$, and $O_5$ as a complete basis for 
the dimension- 7 Higgs-gluon-light quark operators. We can rewrite Eq. \eqref{ldef} as
\begin{equation}
\mathcal L_{\rm eff} = C_1 O_1 + \frac{1}{\Lambda^2} \left( C_3 O_3 + C_4 O_4 + C_5 O_5 \right), \label{ldef1}
\end{equation}
where the re-defined matching coefficients are related to those in Eqs. \eqref{c1hat},\eqref{c2hat}-\eqref{c5hat}, \eqref{c1hatpole}-\eqref{c5hatpole} by,
\begin{align}
C_1 &\equiv \hat C_1 -\frac{m_h^2}{2\Lambda^2} \hat C_2, \label{hatC1} \\
C_3 &\equiv 2 g_s \hat C_2 + \hat C_3, \label{hatC3} \\
C_4 &\equiv \hat C_4, \label{hatC4} \\
C_5 &\equiv 2\hat C_2 + \hat C_5 \label{hatC5} \, .
\end{align}
We will use the basis of Eq. \ref{ldef1} for our phenomenological studies. 

In particular, for SM Higgs production, using $m_t = m_t^{\rm pole}$ in Eq. \eqref{ldef1}, we have
\begin{align}
C_1^{\rm SM,\, pole} (\mu_R) &= \frac{\alpha_s (\mu_R)}{12 \pi v} \left\{ 1 + \frac {\alpha_s (\mu_R)}{4 \pi} \left[ 5 C_A - 3 C_F \right] \right\} + \nonumber\\
&\quad + \frac{7 \alpha_s (\mu_R) m_h^2}{1440 \pi v\, m_t^2} \left\{ 1 + \frac {\alpha_s(\mu_R)}{\pi} \left[ \frac{29}{84} C_A + \frac{19}{21}C_F + \frac 3 2 C_F \ln \left( \frac {m_t^2}{\mu_R^2} \right) \right] \right\}, \label{c1pole} \\
C_3^{\rm SM,\, pole} (\mu_R) &= -\frac{g_s (\mu_R)\alpha_s(\mu_R)}{360 \pi v} \left\{1 + \frac{\alpha_s(\mu_R)}{\pi} \left[ \frac{29}{12} C_A + \frac{25}{3} C_F - 6 \hat C_3^{(1)} -\frac 3 2 C_A \ln \left( \frac {m_t^2}{\mu_R^2} \right) \right] \right\}, \label{c3pole}\\
C_4^{\rm SM,\, pole} (\mu_R) &= \frac{\alpha_s (\mu_R)}{360 \pi v} + \mathcal O \left( \alpha_s^2 (\mu_R) \right), \label{c4pole} \\
C_5^{\rm SM,\, pole} (\mu_R) &= \frac{11 \alpha_s(\mu_R)}{360 \pi v} \left \{1 + \frac{\alpha_s(\mu_R)}{\pi} \left [-\frac {29}{132} C_A + \frac{47}{33} C_F + \frac{18}{11} \hat C_5^{(1)} \right. \right. \nonumber \\
&\qquad \qquad \qquad \qquad  \qquad + \left. \left . \left( -\frac{11}{12}C_A - \frac{2}{3} C_F \right) \ln \left( \frac{m_t^2}{\mu_R^2} \right) \right] \right\}. \label{c5pole}
\end{align}

For the $gg\rightarrow h$ amplitude, $O_3$, $O_4$, and $O_5$ give vanishing contributions at both  tree level and the one-loop level, due either to the lack of quark propagator lines or to the lack of a scale in the diagrams. This leaves us with 
the operator $O_1$ multiplied by the matching coefficient $C_1$ in Eq. \eqref{c1pole} which is defined to include $\mathcal O \left( m_h^2 / m_t^2 \right)$ terms.  This is essentially equivalent to calculating in the $m_t \to \infty$ limit and applying a rescaling factor. For Higgs plus jet production, though, the other operators will come into play and impact differential distributions.

\subsection{Gluon self-interaction}
At $\mathcal O (1/m_t^2)$ in the SM, we also need the dimension-6 gluon self-interaction Lagrangian
which arises from integrating out the top quark and performing Collins-Wilczek-Zee zero-momentum subtraction
to obtain decoupling of the heavy top \cite{Collins:1978wz},
\begin{align}
\mathcal L_{\rm eff}^{\rm SM,self} &= \frac{1}{m_t^2} \left( \frac{g_s \alpha_s}{720 \pi} f_{ABC}G_\nu^{A,\mu} G_\sigma^{B,\nu}G_\mu^{C,\sigma} - \frac{\alpha_s}{60 \pi} D^\sigma G^A_{\sigma\nu}D_\rho G^{A,\rho\nu} \right), \nonumber \\
&\equiv  \frac{1}{m_t^2} \left( \frac{g_s \alpha_s}{720 \pi} \tilde O_3 - \frac{\alpha_s}{60 \pi} \tilde O_4 \right),
\label{Lself}
\end{align}
where the $\tilde O_i$'s are defined to be identical to the $O_i$'s in Eq. \eqref{op2}-\eqref{op5}, but with the Higgs field, $h$, stripped from the operator definition.  Here the matching coefficients are only given at leading order because this is sufficient for NLO Higgs plus jet production.

There is a neat way to obtain the above effective Lagrangian. Using the 
Higgs low-energy theorems \cite{Kniehl:1995tn}, it is easy to see that at leading order matching, the $\mathcal O (1/m_t^2)$ terms in Eq. \eqref{ldef} and \eqref{Lself} can be packaged together in the expression,
\begin{equation}
\mathcal L^{\rm SM} \big|_{\mathcal O(1/m_t^2)} = - \frac{v}{2 m_t^2 \left(1 + \frac h v \right)^2 } \sum_{i=2,3,4,5} \hat C_i \tilde O_i \, .
\label{op-higgs-self}
\end{equation}
 Starting from Eq. \eqref{op-higgs-self}, we use the operator relation of Eq. \eqref{oprel} (which can be applied to $\tilde O_i$'s instead of $O_i$'s by setting $m_h=0$) to eliminate $\tilde O_2$, and further use the relation $\tilde O_4 = \tilde O_5$, valid only at zero-momentum, to eliminate $\tilde O_5$, to reach Eq. \eqref{Lself} which only involves $\tilde O_3$ and $\tilde O_4$.  In a BSM model, the coefficients of the gluon
 self-interactions depend on the nature of the heavy physics which is integrated out.

\section{Lowest Order}
\label{lores}
The lowest order amplitudes for Higgs + jet production including all fermion mass dependence (bottom and top) are given in
Refs. \cite{Baur:1989cm,Ellis:1987xu}.  The effective Lagrangian can be used to obtain the 
contributions from the top quark in the infinite mass approximation, along with the
SM results including
terms of $\mathcal O ( 1 / m_t^2 )$. At the lowest order in $\alpha_s$, $O_3$ is the only dimension-7 operator which contributes to the $gg \rightarrow gh$ channel, while $O_5$ is the only dimension-7 operator which contributes to channels with initial state quarks.
\subsection{Lowest order EFT $q {\bar q} g h$ amplitude}
There are $2$ independent gauge invariant tensor structures for the process $0\rightarrow  q {\bar q} h g $, (where
we consider all momenta outgoing) \cite{Kilgore:2013gba,Gehrmann:2011aa}
\begin{eqnarray} 
{\cal T}_1^\mu &\equiv& i \left( p_{\bar{q}}^\mu {\overline {u}}(p_q)\slashed p_g v(p_{\bar {q}})
-{S_{g{\bar{q}}}\over 2}
{\overline {u}}(p_q)\gamma^\mu v(p_{\bar q}) \right) \label{T1def} \\
{\cal T}_2^\mu &=& i \left( p_{{q}}^\mu {\overline {u}}(p_q)\slashed p_g v(p_{\bar {q}})
-{S_{g{{q}}}\over 2}{\overline {u}}(p_q)\gamma^\mu v(p_{\bar q})\, \right) \label{T2def},
\end{eqnarray}
where $S_{q{\bar{q}}}=(p_q+p_{\bar{q}})^2$, $S_{gq}=(p_g+p_q)^2$,
and $S_{g{\bar{q}}}=(p_g+p_{\bar {q}})^2$.
The $0\rightarrow q {\bar {q}} g h$ amplitude is given in general by,
\begin{eqnarray}
M_{qqgh}^{\alpha,\mu}&=& \Sigma_{i=1,3-5}T^A\biggl(B_1^{\alpha,i}{\cal T}_1^\mu +B_2^{\alpha,i}{\cal T}_2^\mu
\biggr)\, ,
\label{mqg}
\end{eqnarray}
where $\alpha=0,1$ denotes the order of the calculation (LO, NLO), and the sum is over the contributions
of the different operators. 
The tree level amplitude to ${\cal O} (1/\Lambda^2)$ is,
\begin{equation}
M_{qqgh}^{0,\mu}=T^A({\cal T}_1^\mu+{\cal T}_2^\mu)\biggl[C_1\biggl({-4 g_s\over S_{q{\bar q}}}
\biggr)
+{C_5\over \Lambda ^2}(-g_s)\biggr]\, ,
\end{equation}
i.e., the non-vanishing coefficients in Eq. \ref{mqg} are,
\begin{eqnarray}
B_1^{0,1}&=&B_2^{0,1}=C_1\biggl({-4 g_s\over S_{q{\bar q}}}\biggr)
\nonumber \\
B_1^{0,5}&=&B_2^{0,5}={C_5\over \Lambda ^2}(-g_s)\, .
\end{eqnarray}

\subsection{Lowest Order  EFT $gggh$ amplitude}

There are $4$ independent gauge invariant tensor structures for the $0\rightarrow 
g(p_1^\mu)g(p_2^\nu)g(p_3^\rho)h$ 
amplitude \cite{Ellis:1987xu,Gehrmann:2011aa,Kilgore:2013gba},
 assuming
all momenta outgoing and $S_{ij}=2p_i\cdot p_j$,
\begin{eqnarray}
\mathcal{Y}_{0}^{\mu\nu\rho}(p_1,p_2,p_3) & =
&
\left(p_{1}^{\nu}g^{\rho\mu}-p_{1}^{\rho}g^{\mu\nu}\right)\frac{S_{23}}{2}+\left(p_{2}^{\rho}g^{\mu\nu}-p_{2}^{\mu}g^{\nu\rho}\right)\frac{S_{31}}{2}\nonumber \\
 && +\left(p_{3}^{\mu}g^{\nu\rho}-p_{3}^{\nu}g^{\rho\mu}\right)\frac{S_{12}}{2}+p_2^\mu p_{3}^{\nu}p_{1}^{\rho}-p_{3}^{\mu}p_{1}^{\nu}p_{2}^{\rho}\label{eq:y0}\\
\mathcal{Y}_{1}^{\mu\nu\rho}(p_1,p_2,p_3) & = &p_{2}^{\mu}p_{1}^{\nu}p_{1}^{\rho}-p_{2}^{\mu}p_{1}^{\nu}p_{2}^{\rho}\frac{S_{31}}{S_{23}}-\frac{1}{2}p_{1}^{\rho}g^{\mu\nu}S_{12}+\frac{1}{2}p_{2}^{\rho}g^{\mu\nu}\frac{S_{31}S_{12}}{S_{23}} \label{eq:y1} \\
\mathcal{Y}_{2}^{\mu\nu\rho}(p_1,p_2,p_3) &=& \mathcal{Y}_{1}^{\rho\mu\nu}(p_3,p_1,p_2)
\nonumber \label{eq:y2}\\
\mathcal{Y}_{3}^{\mu\nu\rho}(p_1,p_2,p_3)
 &=& \mathcal{Y}_{1}^{\nu\rho,\mu}(p_2,p_3,p_1)\, . \nonumber \label{eq:y3}
\end{eqnarray}
An arbitrary $gggh$ amplitude is written as
\begin{eqnarray}
{\cal M}_{gggh}^{\alpha,\mu\nu\rho}&=&f_{ABC}\Sigma_i
\biggl\{A_{0}^{\alpha,i}(p_1,p_2,p_3)\mathcal{Y}_{0}^{\mu\nu\rho}(p_1,p_2,p_3)+
\nonumber \\ &&
\sum_{m=1,2,3}A_{m}^{\alpha,i}(p_1,p_2,p_3)\mathcal{Y}_{m}^{\mu\nu\rho}(p_1,p_2,p_3) 
\biggr\}\, ,
\label{eq:sum-MiYi}
\end{eqnarray}
where again $\alpha$ = 0,1 for the lowest order (LO) and next-to-leading order (NLO) contributions,  $i$ is the contribution
corresponding to $O_i$, and
\begin{eqnarray}
A_2^{\alpha,i}(p_1,p_2,p_3)&=& A_1^{\alpha,i}(p_3,p_1,p_2)
\nonumber \\
A_3^{\alpha,i}(p_1,p_2,p_3)&=& A_1^{\alpha,i}(p_2,p_3,p_1)\, .
\end{eqnarray}

The LO contributions from $O_{1}$ and $O_{3}$ are
\begin{eqnarray}
A_0^{0,1}(p_1,p_2,p_3) & =&8g_s C_1\left(\frac{1}{S_{12}}+\frac{1}{S_{23}}+\frac{1}{S_{31}}\right)
\nonumber \\
A_1^{0,1}(p_1,p_2,p_3) & =&\frac{8g_s C_1}{S_{31}}
\nonumber \\
A_0^{0,3}(p_1,p_2,p_3) & = &{C_3\over \Lambda^2}6
\nonumber \\
A_1^{0,3}(p_1,p_2,p_3) & =&0
\, ,
\end{eqnarray}
while the $O_{5}$ contribution vanishes.

\subsection{Squared amplitudes}
To obtain squared amplitudes, we need the interference between the Lorentz / Dirac tensor structures, and the interference between the color structures. For the $qg \rightarrow qh$ squared amplitude, 
the interferences between the tensor structures are (omitting the ones which can be obtained from $q \leftrightarrow \bar q$ crossing symmetry between ${\cal T}_1$ and ${\cal T}_2$).
\begin{align}
\sum_A \rm{tr} \left( T^A T^A \right) &= \frac{N_c^2 - 1}{2}, \\
- \sum_{\rm spins} \mathcal T_1^\mu \, \mathcal T_{1,\mu}^\dagger &= -(1-\epsilon) S_{q \bar q} S_{gq}^2, \\
- \sum_{\rm spins} \mathcal T_1^\mu \mathcal T_{2,\,u}^\dagger &= -\epsilon S_{q\bar q}S_{gq}S_{g\bar q}\, ,
\end{align}
where external fermion spinors are implicit and we work in $N=4-2\epsilon$ dimensions.
The $q\bar q \rightarrow gh$ squared amplitude can be obtained from crossing the $qg \rightarrow qh$ squared amplitude. For the $gg\rightarrow gh$
squared amplitude, the interferences between the tensor structures are,
\begin{align}
\sum_{ABC} f^{ABC} f^{ABC} &= N_c (N_c^2 - 1), \\
-\sum_{\rm spins} \mathcal Y_0^{\mu \nu \rho} \mathcal Y^\dagger_{0,\mu \nu \rho} &= \left( 1 - \frac 3 2 \epsilon \right) S_{12} S_{23} S_{31}, \\
-\sum_{\rm spins} \mathcal Y_1^{\mu \nu \rho} \mathcal Y^\dagger_{0,\mu \nu \rho} &= \frac 1 2 (1 - \epsilon) S_{12}^2 S_{31}, \\
-\sum_{\rm spins} \mathcal Y_1^{\mu \nu \rho} \mathcal Y^\dagger_{1,\mu \nu \rho} &= \frac 1 2 (1 - \epsilon) \frac{S_{12}^3 S_{31}}{S_{23}}, \\
-\sum_{\rm spins} \mathcal Y_1^{\mu \nu \rho} \mathcal Y^\dagger_{2,\mu \nu \rho} &= \frac 1 4 S_{12} S_{31}^2\, ,
\end{align}
where we have omitted terms which can be obtained from cyclic permutations.

Here we present squared amplitudes, summed (but not averaged) over initial and final state spins, with $\mathcal O(\epsilon)$ terms omitted. For $gg \rightarrow gh$, the squared amplitude from the $O_1$ operator is \cite{Ellis:1987xu}
\begin{equation}
\sum_{\rm spins} \left| M_{gg \rightarrow gh,\,O_1}^{(0)} \right|^2 = 384\, C_1^2\, \frac{m_h^8 + s^4 + t^4 + u^4}{s t u},
\label{gg2gh-LO-O1}
\end{equation}
while the $O_1$-$O_3$ interference contribution is
\begin{equation}
\sum_{\rm spins} M_{gg \rightarrow gh,\,O_1}^{(0)} \cdot M_{gg \rightarrow gh,\,O_3}^{(0),\dagger} + {\rm c.c.} = 1152\, C_1 C_3\, \frac{m_h^4}{\Lambda^2}.
\label{gg2gh-LO-O3}
\end{equation}
Interestingly, the $O_1$ contribution, Eq. \eqref{gg2gh-LO-O1}, corresponding to a rescaled $m_t \to \infty$ approximation, grows as $p_T^2$ for high $p_T$ Higgs production, while the $O_1$-$O_3$ interference contribution, Eq. \eqref{gg2gh-LO-O3}, remains constant and therefore diminishes in relative importance, in contrary to the generic behavior of higher-dimensional operators. This results in suppressed top mass dependence in Higgs differential distributions in the gluon channel, and will be explained by the helicity structure of the amplitudes in the soft Higgs limit, i.e. the limit $m_h^2 < (p_T^2, s, -t, -u)$, discussed in Section \ref{nloreal}.

For $qg \to qh$, the squared amplitude from the $O_1$ operator is \cite{Ellis:1987xu}
\begin{equation}
\sum_{\rm spins} \left| M_{qg \rightarrow qh,\,O_1}^{(0)} \right|^2 = 64\, C_1^2 \, \frac{s^2+u^2}{-t},
\end{equation}
while the $O_1$-$O_5$ interference contribution is
\begin{equation}
\sum_{\rm spins} M_{qg \rightarrow qh,\,O_1}^{(0)} \cdot M_{qg \rightarrow qh,\,O_5}^{(0),\dagger} + {\rm c.c.} = -32\, C_1 C_5\, \frac{ s^2 + u^2 }{\Lambda^2}
\label{gg2gh-LO-O5}
\end{equation}
The results, crossed into the $q\bar q \rightarrow gh$ channel, are
\begin{align}
\sum_{\rm spins} \left| M_{qg \rightarrow qh,\,O_1}^{(0)} \right|^2 &= 64\, C_1^2 \, \frac{t^2+u^2}{s}, \\
\sum_{\rm spins} M_{qg \rightarrow qh,\,O_1}^{(0)} \cdot M_{qg \rightarrow qh,\,O_5}^{(0),\dagger} + {\rm c.c.} &= 32\, C_1 C_5\, \frac{ t^2 + u^2 }{\Lambda^2}.
\end{align}

\section{Renormalization of dimension-7 operators}
\label{rencond}
In this section, we use the basis $ O_6 \cong m_h^2 O_1$, $O_3$, $O_4$, and $O_5$, described in Section II.B, for the dimension-7 operators. 
In addition to the renormalization of the QCD coupling constant and self energies in both QCD vertices and the $O_i$ operators, we need to renormalize the $C_i$ matching coefficients. The renormalization of $C_1$ is well known \cite{KlubergStern:1974rs,Tarrach:1981bi,Grinstein:1988wz}, and is identical to the renormalization of $\alpha_s$ at one-loop. The renormalization of $C_3$ and $C_5$ are different, and they will be presented as the sum of $\alpha_s$ renormalization and an extra piece. The renormalization of $C_3$ was found in Ref. \cite{Gracey:2002he}. The renormalization of $C_5$ is a new result.

The unrenormalized effective Lagrangian coupling the Standard Model Higgs boson to gluons is,
\begin{equation}
\mathcal L_{\rm eff}=C_1^{\rm bare}O_1^{\rm bare}+\Sigma_{i=3-5} {C_i^{\rm bare} \over{ \Lambda^2}} O_i^{\rm bare}\,,
\end{equation}
where $\Lambda$ is a constant power expansion parameter that should not depend on $\mu_R$, so in this section we will allow $\Lambda$ to be equal to the top quark pole mass in the case of SM Higgs production, but not the running $\overline{\rm MS}$ mass. The operators $O_i^{\rm bare}$ are  defined in the same way as $O_i$, but with all the fields and couplings replaced by bare quantities.
$O_4$ is needed only at LO, so we will not discuss its one-loop renormalization. In our operator basis, the one-loop mixing matrix is diagonal, so we can write
\begin{align}
C_i^{\rm bare} &= C_i + \delta C_i = Z_i C_i = (1 + \delta Z_i) C_i\, .
\end{align}
The renormalization constants $Z_i$ are found using two different methods. The first one is to calculate one-loop $ggh$, $gggh$, and $q\bar{q}gh$ amplitudes on-shell, and impose transverse gluon polarizations to eliminate spurious mixing into gauge non-invariant operators. The second method is to calculate these one-loop amplitudes off-shell to reduce the number of diagrams needed, and use the background field method \cite{Abbott:1980hw}  to preserve gauge-invariance.  In either method, the divergences are matched to the tensor structures arising from the various operators in order to extract the renormalization of the $C_i$.
The renormalization counterterms are given by,
\begin{align}
\delta Z_1 &= \delta Z_{\alpha_s}, \label{deltaZ1} \\
\delta Z_3 &= \frac{3}{2} \delta Z_{\alpha_s} + \frac{\alpha_s} {2 \pi \epsilon} (4\pi)^\epsilon\, r_\Gamma \, 3 C_A ,\label{deltaZ3} \\
\delta Z_5 &= \delta Z_{\alpha_s} + \frac{\alpha_s} {2 \pi \epsilon} (4\pi)^\epsilon\, r_\Gamma \left( \frac{11}{6}C_A + \frac{4}{3} C_F \right),\label{deltaZ5}
\end{align}
where $r_\Gamma$ is given in Eq. \eqref{rGamma}, and
\begin{align}
\delta Z_{\alpha_s} &= \frac{\alpha_s}{\pi \epsilon} (4\pi)^\epsilon r_\Gamma\, b_0, \\
b_0 &= \left( \frac{11}{12} C_A - \frac{1}{6} n_{lf} \right)\, ,
\end{align}
is the one-loop renormalization factor for the strong coupling $\alpha_s$ in an $n_{lf}=5$ flavor
theory, proportional to the beta function.

By using
\begin{equation}
\frac{d \ln C_i}{d \ln \mu_R} = -\frac {d \ln Z_i}{d \ln \mu_R},
\end{equation}
we have the following renormalization group running equations,
\begin{align}
\frac {d} {d \ln \mu_R} \ln \left( \frac {C_1}{g_s^2} \right) &= \mathcal O(\alpha_s^2(\mu_R)), \label{c1run} \\
\frac {d} {d \ln \mu_R} \ln \left( \frac {C_3}{g_s^2} \right) &= \frac{\alpha_s(\mu_R)}{\pi}\, 3 C_A, \label{c3run} \\
\frac {d} {d \ln \mu_R} \ln \left( \frac {C_5}{g_s^2} \right) &= \frac{\alpha_s(\mu_R)}{\pi} \left( \frac {11}{6}C_A + \frac{4}{3} C_F \right) \label{c5run}.
\end{align}
The leading-logarithmic solutions to the renormalization group  running  of Eqs. \eqref{c1run}-\eqref{c5run} are
\begin{align}
C_1(\mu_R) / g_s^2 (\mu_R) &= C_1(\mu_0) / g_s^2 (\mu_0), \\
C_3(\mu_R) / g_s^3 (\mu_R) &= \left( \frac{\alpha_s(\mu_R)}{\alpha_s(\mu_0)} \right)^{-\frac {3 C_A}{2b_0}} \cdot C_3(\mu_0) / g_s^3 (\mu_0), \\
C_5(\mu_R) / g_s^2 (\mu_R) &= \left( \frac{\alpha_s(\mu_R)}{\alpha_s(\mu_0)} \right)^{-\frac {1}{2b_0} \left( \frac{11}{6} C_A + \frac 4 3 C_F \right)} \cdot C_5(\mu_0) / g_s^2 (\mu_0),
\end{align}
which in principle allows us to perform matching at the new physics scale $\Lambda$, and use renormalization group running to obtain $C_i$ at $\mu_R \sim m_h$, hence resumming large logarithms of $\Lambda/ \ m_h$.

\section{NLO virtual corrections}
\label{lhcres}
\subsection{Methods}
All our NLO calculations are done using $O_1$, $O_3$, and $O_5$ as a basis of operators, as described in Section \ref{subsec:alternativeBasis}, with  $O(m_h^2 / m_t^2)$ terms included in the $C_1$ matching coefficient to absorb the dimension-7 operator $O_6$ operator in Eq. \eqref{opRelation}. When calculating NLO virtual amplitudes for $O_5$, we exploit equations of motions to use the $O_5'$ operator in Eq. \eqref{op5prime} instead. The NLO virtual diagrams needed for $O_1$ are also the only ones needed for $O_3$ and $O_5'$. Our amplitude-level results, given as coefficients for the tensor structures in Eqs. \eqref{T1def},\eqref{T2def},\eqref{eq:y0}-\eqref{eq:y3}, are valid in both the conventional dimensional regularization (CDR) scheme in $D$ dimensions and the t'Hooft-Veltman scheme which has loop momenta in $D$ dimensions and external leg momenta in $4$ dimensions.

The one-loop virtual calculation is done as follows. The software FeynRules \cite{Alloul:2013bka} is used to generate Feynman rules for each of the operators. FeynArts \cite{Hahn:2000kx} is used to generate Feynman diagrams and produce expressions for the amplitudes by using the Feynman rules, with loop integrations unperformed. FormCalc \cite{Hahn:1998yk} is used to perform the numerator algebra and loop integration, producing results in terms of one-loop tensor integrals (up to rank-5 box integrals). The tensor integrals are subsequently reduced to scalar integrals in $D$ dimensions using FeynCalc \cite{Mertig:1990an}, and combined with the explicit results for the scalar integrals \cite{Ellis:2007qk} to produce our final analytic results for the one-loop virtual amplitudes. Alternatively, the tensor integrals can be evaluated numerically using LoopTools \cite{Hahn:1998yk} without analytic reduction to scalar integrals, and we have checked that the results agree numerically with our analytic formulas for the one-loop amplitudes.\footnote{We find that there are some special tensor integrals which cannot be reduced to scalar integrals correctly by FeynCalc in $D$ dimensions, but this problem has not affected our calculation, since the end results are in agreement with LoopTools.}

\subsection{One loop $q\bar qgh$ amplitudes}
The one-loop virtual amplitudes for $0 \to q {\bar q}g h$ and the real emission amplitudes for $0 \rightarrow q\bar q ggh$ are responsible for both $qg \to h+j+X$ and the $q \bar q \to h+j+X$, where $j=g,q$ or ${\bar q}$.

We list  only the $B_2$ contributions for the virtual one-loop diagrams
from each of the operators since $B_1$ can be obtained by exchanging $S_{gq}$ and $S_{g{\bar q}}$.
 The 
virtual contribution proportional to $C_4$ vanishes.

The non-vanishing one-loop coefficients, $B_2^{1,i}$ defined in Eq. \ref{mqg}, from the operators $O_i$ are,
\begin{eqnarray}
B_2^{1,1}&=& {\alpha_s(\mu_R)\over 4\pi} r_\Gamma\biggl({4\pi\mu_R^2\over m_h^2}
\biggr)^\epsilon
B_2^{0,1}\biggl[ N_c V_1+{1\over N_c}V_2+n_{lf}V_3\biggr]
\nonumber \\
B_2^{1,3}&=&{C_3\over m_t^2} {\alpha_s(\mu_R)\over 8\pi}N_c
\nonumber \\
B_2^{1,5}&=&
 {\alpha_s(\mu_R)\over 4\pi} r_\Gamma\biggl({4\pi\mu^2\over m_h^2}
 \biggr)^\epsilon
B_2^{0,5}\biggl[ N_c W_1+{1\over N_c}W_2+n_{lf}W_3\biggr]\, ,
\label{b2def}
\end{eqnarray}
where
\begin{equation}
r_\Gamma\equiv {\Gamma^2(1-\epsilon)\Gamma(1+\epsilon)\over \Gamma(1-2\epsilon)}\, .
\label{rGamma}
\end{equation}
Analytic expressions for the functions $V_i$ and $W_i$ are given in Appendix \ref{sec:appa}. 

The $0\rightarrow q\bar q gh$ amplitude involves one ordinary QCD coupling and one EFT coupling, both of which need counterterms. The sum of the counterterms is
\begin{eqnarray}
M^{CT,\mu}_{q\bar qgh}&=& 
{3\over 2} \delta Z_{\alpha_s} M_{q\bar qgh}^{0,\mu}
 -g_s(\mu_R)
 T^A({\cal T}_1+{\cal T}_2)^\mu {\alpha_s(\mu_R)\over 2 \pi\epsilon} \left( \frac {11}{6} C_A + \frac 4 3 C_F \right)  {C_5\over \Lambda^2},
\label{qgren}
\end{eqnarray}
where the renormalization for the $O_1$ amplitude is simply proportional to 3 times the $g_s$ renormalization \cite{Schmidt:1997wr,Ravindran:2002dc}, whereas there is an extra term for the $O_5$ amplitude because the $C_5$ renormalization in Eq. \eqref{deltaZ5} is not proportional to $\delta Z_{\alpha_s}$.

The renormalized one-loop virtual amplitude is then,
\begin{equation}
M_{q\bar qgh}^{V+CT,^\mu}=\biggl({4\pi\mu_R^2\over m_h^2}\biggr)^\epsilon r_\Gamma \biggl\{
\biggl[{A_{V2}\over\epsilon^2}
+{A_{V1}\over\epsilon}\biggr]M_{q\bar qgh}^\mu+
\biggl({\cal T}_1+{\cal T}_2\biggr)^\mu T^A A_{V0} \biggr\}\, ,
\label{virt_qg}
\end{equation}
where
\begin{eqnarray}
A_{V2}&=&{\alpha_s(\mu_R)\over 4\pi}\biggl(-2N_c+{1\over N_c}\biggr)\nonumber \\
A_{V1}&=&{\alpha_s(\mu_R)\over 4\pi}
\biggl\{ N_c \ln\biggl({-S_{gq} \over m_h^2}\biggr)
+  N_c \ln\biggl({-S_{g\bar q} \over m_h^2}\biggr)
-{1\over N_c}\ln\biggl({-S_{q{\bar q}}\over m_h^2}\biggr)
\biggr]\, .
\label{virts}
\end{eqnarray}
Note that the finite contribution to the virtual amplitude, $A_{V0}$, is not proportional to the LO result.  $A_{V0}$ 
is just the contribution from the finite terms in defined in Eq. \ref{b2def} and Appendix \ref{sec:appa}.

\subsection{One loop $gggh$ amplitudes}

The $1$-loop virtual results are,
\begin{eqnarray}
A_0^{1,1} & =&
\frac{\alpha_{s}(\mu_R)}{4\pi}r_{\Gamma}\left(\frac{4\pi\mu^{2}}{m_{h}^{2}}\right)^{\epsilon}N_{c}\, U_{1}\, 
A_0^{0,1}
\nonumber \\
A_1^{1,1} & =&
\frac{\alpha_{s}(\mu_R)}{4\pi}r_{\Gamma}\left(\frac{4\pi\mu^{2}}
{m_{h}^{2}}\right)^{\epsilon}\left[N_{c}\, U_{1}\, A_1^{0,1}+\frac{8g_{s}\left(N_{c}-N_{lf}\right)S_{23}}{3S_{12}^{2}}\right]
\nonumber \\
A_0^{1,3} & =&\frac{\alpha_{s}(\mu_R)}{4\pi}r_{\Gamma}\left(\frac{4\pi\mu^{2}}
{m_{h}^{2}}\right)^{\epsilon}N_{c}\, U_{3}\, A_0^{0,3}
\nonumber \\
A_1^{1,3}&=&0
\nonumber \\
A_0^{1,5}&= & 0
\nonumber \\
A_1^{1,5}&= & - \frac{g_{s}\alpha_{s}(\mu_R)}{4\pi}\cdot\frac{2S_{23}}{3S_{12}}
\, .
\nonumber \\
\end{eqnarray}
Analytic expressions for the functions $U_1$ and $U_3$ are given in Appendix \ref{sec:appa}. 

The counterterm from renormalization for the QCD coupling and the EFT matching coefficients is,
\begin{eqnarray}
{\cal M}_{gggh}^{CT,\mu\nu\rho}&=&f_{ABC}\biggl\{\biggl(
\delta Z_1+{1\over 2}\delta Z_{\alpha_s} \biggr)
\biggl(
A_{0}^{0,1}(p_1,p_2,p_3)\mathcal{Y}_{0}^{\mu\nu\rho}(p_1,p_2,p_3)+
\nonumber \\ &&
\sum_{m=1,2,3}A_{m}^{0,1}(p_1,p_2,p_3)\mathcal{Y}_{m}^{\mu\nu\rho}(p_1,p_2,p_3) 
\biggl)
\nonumber \\ &&
+\delta Z_3
\biggl(
A_{0}^{0,3}(p_1,p_2,p_3)\mathcal{Y}_{0}^{\mu\nu\rho}(p_1,p_2,p_3)+
\nonumber \\ &&
\sum_{m=1,2,3}A_{m}^{0,3}(p_1,p_2,p_3)\mathcal{Y}_{m}^{\mu\nu\rho}(p_1,p_2,p_3) 
\biggl)
\biggr\}
\label{eq:sum-MiYi}
\end{eqnarray}

\subsection{Soft and Collinear real contributions}
\subsubsection{Soft - $qg$ channel}
We combine the virtual and real amplitudes using the  2 cut-off phase space slicing method to regulate the soft and collinear singularities in $D$ dimensions \cite{Harris:2001sx} for the $qg\rightarrow h+j+X$ and $gg\rightarrow h+j+X$ channels. The results for $q\bar q\rightarrow h+j+X$ can be obtained in a similar manner and are included in our
 numerical results. 

To find the NLO cross section, we integrate the LO, NLO virtual, soft and collinear contributions over the 2-body final state phase space, and integrate the hard non-collinear contribution over the 3-body final phase space.  The total answer is finite and independent of $\delta_c$ and $\delta_s$.

The soft contribution is defined as the contribution from real gluon emission, $qg\rightarrow q g h$,
where the outgoing gluon has an energy less than a small cut-off \cite{Harris:2001sx},
\begin{equation}
 E_g < \delta_s{\sqrt{s}\over 2}\, .
 \end{equation}
where $\delta_s$ is an arbitrary small number. For the $qg$ initial state, $s=S_{g\bar q}$, $t=S_{q\bar q}$, and $u=S_{gq}$.

The soft contribution is found by integrating the eikonal approximation to the $qg\rightarrow qh + g_{\rm soft}$ 
amplitude-squared and integrating
over the soft gluon phase space following exactly the procedure of Ref. \cite{Harris:2001sx}.
The required integrals are found in Ref. \cite{Beenakker:1988bq}.
The soft result is, 
\begin{equation}
\mid M_{qg \rightarrow qh}^{\rm soft}\mid^2=-{\alpha_s(\mu_R)\over 4\pi}
r_\Gamma
\biggl({4\pi \mu_R^2 \over m_h^2 }\biggr)^\epsilon\mid {M}_{qg \rightarrow qh}^{(0)}\mid^2 
\biggl\{
A_{2S}{1\over \epsilon^2}+A_{1S}{1\over\epsilon}+A_{0S}\biggr\}\, ,
\label{qg_softans}
\end{equation}
where,
\begin{align}
A_{S2}&=-{34\over 3}, \nonumber \\
A_{S1}&= \frac {68}{3} \ln\delta_s-
6 \ln\biggl({ m_h^2 \beta_H\over -u}\biggr) - 6\ln \biggl( {m_h^2 \over s} \biggr)
+ {2\over 3} \ln\biggl({ m_h^2 \beta_H\over -t}\biggr) \nonumber \\
&-
\ln\biggl(
{s \over m_h^2}\biggr)A_{S2}\nonumber \\
A_{S0}&= -\frac{68}{3} \ln^2 \delta_s + 12 \left(\ln \frac{m_h^2 \beta_h}{-u} \right) \ln \delta_s + 12 \ln \left( \frac{m_h^2}{s} \right) \ln \delta_s - \frac 4 3 \ln \left( \frac{m_h^2 \beta_h }{-t} \right) \ln \delta_s \nonumber \\
& - 3 \ln^2 \left( \frac{m_h^2}{s} \right) - 3 \ln^2 \left( \frac{m_h^2 \beta_h}{-u} \right) + \frac 1 3 \ln^2 \left( \frac {m_h^2 \beta_h}{-t} \right) \nonumber \\
& \quad + \left[ \ln^2 \left( \frac{s}{m_h^2} \right) - \frac{\pi^2}{3} \right] \frac{A_{S2}}{2}\, ,
\end{align}
and  $\beta_H=1-m_h^2 /s$.   

The hard contribution to the real gluon emission
process $qg\rightarrow qgh$  contains collinear singularities,
\begin{equation}
{\sigma}_{\rm real}={\sigma}_{\rm hard/collinear}+{\sigma}_{\rm hard/non-collinear}\, .
\end{equation}
The hard/non-collinear terms  arising from $i\rightarrow j$ parton splitting  are finite and satisfy,
\begin{eqnarray}
E_g &> &\delta_s
{\sqrt{s}\over 2}\nonumber \\
\mid S_{ij}\mid & >& \delta_c s \, ,
\end{eqnarray}
where $\delta_c$ is an arbitrary collinear cut-off and is typically $\ll \delta_s$.
These terms can be integrated numerically using the amplitudes given in Appendix \ref{sec:appb}.

\subsubsection{Final State Collinear - $qg$ channel}
The hard collinear contribution to the partonic
cross section from $q\rightarrow q g$ splitting in the final state is \cite{Harris:2001sx},
\begin{eqnarray}
 {\hat \sigma}_{qg\rightarrow qgh}^{HC,f}&=&
{\hat\sigma}^{LO}_{qg}{\alpha_s(\mu_R)\over 2 \pi} r_\Gamma
\biggl({4\pi\mu_R^2\over s}\biggr)^\epsilon\biggl\{ \biggl( {1\over\epsilon}-\ln \delta_c\biggr)
C_F\biggl[2\ln \biggl({\delta_s\over \beta_H}\biggr)  +{3\over 2}\biggr]
\nonumber \\
 &&
-{\pi^2\over 3}-
\ln^2\biggl({\delta_s\over \beta_H}\biggr)+{7\over 2}\biggr\}\, .
\label{hard_fqg}
\end{eqnarray}

\subsubsection{Soft - $gg$ channel}
The contribution from soft gluon emission results from integrating the eikonal approximation
to the $gg\rightarrow gh + g_{\rm soft}$ matrix-element squared
 over the soft gluon phase space and yields,
\begin{eqnarray}
\mid M^{soft}_{gg\rightarrow gh}\mid^2
&=&
{\alpha_s(\mu_R)\over \pi}
r_\Gamma
\biggl({4\pi\mu_R^2\over m_h^2}\biggr)^\epsilon
\biggl\{
{A_{g2}\over \epsilon^2}+{A_{g1}\over\epsilon}+A_{g0}\biggr\}\mid { M}_{gg\rightarrow gh}^{(0)}\mid^2,
\end{eqnarray}
with
\begin{eqnarray}
A_{g2}&=& {3\over 2}N_c={9\over 2},
\nonumber \\
A_{g1}&=& {N_c\over 2}\biggl\{ -6\log(\delta_s)
+\ln\biggl({m_h^2\over S_{12}}\biggr)
+\ln\biggl({m_h^2\beta_H\over- S_{13}}\biggr)
+\ln\biggl({m_h^2\beta_H \over -S_{23}}\biggr)\biggr\}
\nonumber \\
&-&
\ln\biggl(
{S_{12}\over m_h^2}\biggr)A_{g2},\nonumber \\
A_{g0}&=&{N_c\over 4}
\biggl\{
 12\ln^2(\delta_s)+
 \ln^2\biggl({m_h^2\over S_{12}}\biggr)
+\ln^2\biggl({m_h^2\beta_H\over- S_{13}}\biggr)
+\ln^2\biggl({m_h^2\beta_H \over -S_{23}}\biggr)
\nonumber \\
&&-4\ln\delta_s\biggl[\ln\biggl({m_h^2\over S_{12}}\biggr)
+\ln\biggl({m_h^2\beta_H\over- S_{13}}\biggr)
+\ln\biggl({m_h^2\beta_H \over -S_{23}}\biggr)\biggr]
\nonumber \\ &&
+2Li_2\biggl({-S_{23}\over S_{12}\beta_H}\biggr)+
2Li_2\biggl({-S_{13}\over S_{12}\beta_H}\biggr)\biggr\} \nonumber \\
&& + \left[ \ln^2 \left( \frac {S_{12}}{m_h^2} \right) - \frac {\pi^2}{3} \right] \frac{A_{g2}} 2.
\end{eqnarray}

\subsubsection{Final State Collinear - $gg$ channel}
The hard collinear contributions from gluon splitting in the final state are \cite{Harris:2001sx} ,
\begin{eqnarray}
{\hat \sigma}^{HC,f}_{gg\rightarrow ggh} &=&
{\hat\sigma}^{LO}_{gg\rightarrow gh}{\alpha_s(\mu_R)\over 2 \pi} 
r_\Gamma
\biggl({4\pi\mu_R^2\over s}\biggr)^\epsilon N_c\biggl\{
\biggl({1\over\epsilon}-\ln \delta_c\biggr)
\biggl[2\ln \biggl({\delta_s\over \beta_H}\biggr)  +{11\over 6}\biggr]
\nonumber \\
&&-{\pi^2\over 3}-
\ln^2\biggl({\delta_s\over \beta_H}\biggr)+{67\over 18}\biggr\}\, , \\
{\hat \sigma}^{HC,f}_{gg\rightarrow q\bar qh} &=&
{\hat\sigma}^{LO}_{gg\rightarrow gh}{\alpha_s(\mu_R)\over 2 \pi} 
r_\Gamma
\biggl({4\pi\mu_R^2\over s}\biggr)^\epsilon n_{lf} \biggl\{
\biggl({1\over\epsilon}-\ln \delta_c\biggr)
\biggl( -{1 \over 3} \biggr)
-{5 \over 9} \biggr\}\, ,
\end{eqnarray}

\subsubsection{Initial State Collinear - all channels}
The contribution from collinear splitting in the initial state is combined with the renormalization of the PDFs to
obtain the result given in \cite{Harris:2001sx}, applicable to all channels,
\begin{align}
d \hat \sigma_{1+B \rightarrow 3+4+5}^{\rm initial+PDF} &= d \hat \sigma^{\rm LO}_{1+2' \rightarrow 3+4} \frac{\alpha_s (\mu_R)}{2\pi} \frac{\Gamma(1-\epsilon)}{\Gamma(1-2\epsilon)} \left[ \left( \frac{4\pi \mu_R^2}{s} \right)^\epsilon  \tilde f_{2'/B} (z, \mu_F) \right. \nonumber \\
&\quad + \left. \frac 1 \epsilon \left( \frac{4\pi \mu_R^2}{\mu_F^2} \right)^\epsilon A_1^{sc}(2 \rightarrow 2'+5)\, f_{2/B} (z,\mu_F) \right],
\end{align}
where the initial state hadron $B$ splits into a parton $2'$ which scatters with the initial state parton $1$ and a parton $5$ which goes into the final state. The redefined parton distribution function $\tilde f$ is given by \cite{Harris:2001sx}
\begin{align}
\tilde f_{c/B} ( x, \mu_f ) &= \sum_{c'} \int_x^{1 - \delta_s \delta_{c c'} } \frac {dy}{y} f_{c'/B} (x/y, \mu_f) \tilde P_{c c'} (y), \\
\tilde P_{ij} (y) &= P_{ij} (y) \ln \left( \delta_c \frac{1-y}{y} \frac{s}{\mu_f^2} \right) - P_{ij}' (y),
\end{align}
where $P_{ij}$ and $P_{ij}'$ are the $\mathcal O(\epsilon^0)$ and $\mathcal O(\epsilon)$ parts of the $D$-dimensional splitting function. The soft-collinear term $A_1^{sc}$, from the soft cutoff on initial state gluon emission, is given by \cite{Harris:2001sx}
\begin{align}
A_1^{sc} ( q \rightarrow qg ) &= C_F ( 2 \ln \delta_s + 3 / 2), \\
A_1^{sc} ( g \rightarrow gg ) &= 2 C_A \ln \delta_s + (11 C_A - 2 n_{lf} ) / 6, \\
A_1^{sc} ( g \rightarrow q\bar q) &= 0.
\end{align}

\subsection{Higher-dimensional gluon self interaction contribution}
\label{sec:self}
\begin{figure}
\begin{centering}
\includegraphics[scale=0.6]{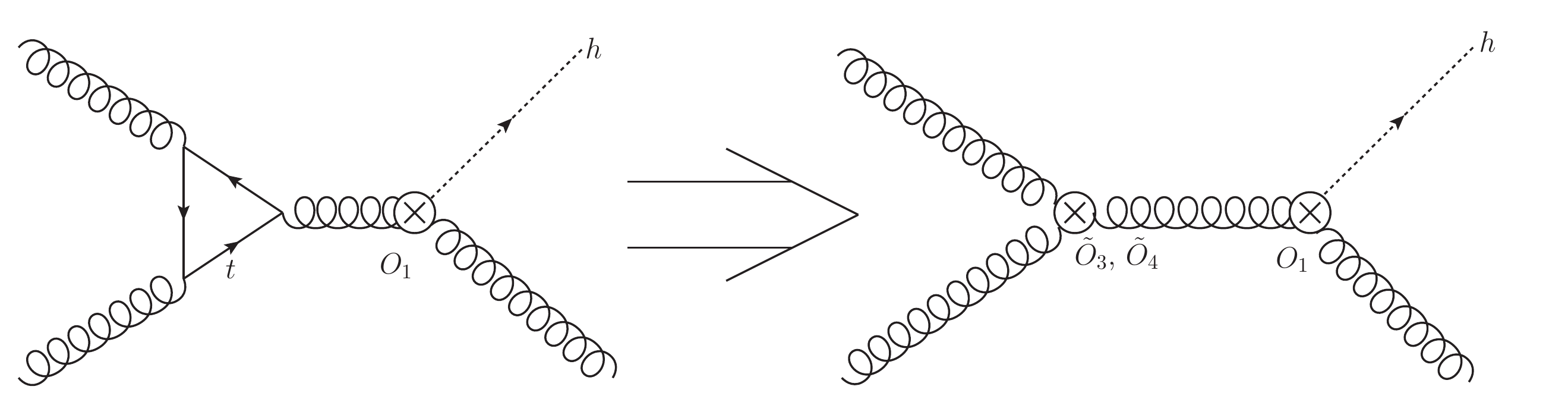}
\par\end{centering}
\caption{An example diagram showing the  $\mathcal O(1/m_t^2)$ gluon self-interaction vertex from integrating out the top quark. The Higgs is produced through the $O_1$ operator in the $m_t \to \infty$ limit, but the overall power of this Feynman diagram is still of $\mathcal O(1/m_t^2)$ and should be considered on the same footing as diagrams producing the Higgs through $1/m_t^2$-suppressed dimension-7 operators.}
\label{fig:self}
\end{figure}
In Fig. \ref{fig:self} we give an example Feynman diagram which involves Higgs coupling in the $m_t \to \infty$ limit but contains an $\mathcal O(1/m_t^2)$ gluon-self coupling EFT vertex. Other diagrams of this type involve top quark loops as self-energy corrections of internal gluon propagators. These diagrams can be trivially calculated exactly, but we choose to use the EFT Lagrangian in Eq. \eqref{Lself} which gives the expansion to $\mathcal O(1/m_t^2)$. The contributions of these diagrams are of NLO order in $\alpha_s$ counting and $\mathcal O(1/m_t^2)$ in EFT power counting.

The contribution to the $0 \rightarrow q\bar q gh$ amplitude is
\begin{equation}
-8 g_s^3 \frac{1}{m_t^2} \tilde C_4 (\mathcal T_1 + \mathcal T_2) T^A = \frac{g_s^5}{30\pi^2 m_t^2} (T_1 + T_2) T^A,
\end{equation}
while the contribution to the $0 \rightarrow gggh$ amplitude is
\begin{align}
 & \quad24g_{s}^{2}f^{ABC}\tilde{C}_{3}\left(\frac{S_{23}}{S_{12}}\mathcal{Y}_{1}^{\mu\nu\rho}+\frac{S_{12}}{S_{31}}\mathcal{Y}_{2}^{\mu\nu\rho}+\frac{S_{31}}{S_{23}}\mathcal{Y}_{3}^{\mu\nu\rho}\right)\\
 & =\frac{1}{120\pi^{2}}g_{s}^{5}f^{ABC}\left(\frac{S_{23}}{S_{12}}\mathcal{Y}_{1}^{\mu\nu\rho}+\frac{S_{12}}{S_{31}}\mathcal{Y}_{2}^{\mu\nu\rho}+\frac{S_{31}}{S_{23}}\mathcal{Y}_{3}^{\mu\nu\rho}\right),
\end{align}
where the $T_i$ and $Y_i$ tensor structures are given in Eqs. \eqref{T1def},\eqref{T2def},\eqref{eq:y0}-\eqref{eq:y3}.

\section{NLO real emission helicity amplitudes}
The helicity amplitudes for the production of Higgs plus two jets in the $m_t \to \infty$ limit, i.e. the $O_1$ contribution, was worked out long ago \cite{Dawson:1991au,Kauffman:1996ix}. We will calculate the amplitudes for dimension-7 operators.  The all-gluon amplitudes will be given in this section, while amplitudes involving quarks will be given in Appendix \ref{sec:appb}.  The $O_4$ and $O_5$ operators, which involve quark bilinears, do not contribute to tree amplitudes without external quark legs, so only $O_1$ and $O_3$ will appear here.

Amplitudes for the $G^3$ operator without the Higgs, as a model for higher-dimensional modifications of the SM QCD sector, were studied in Refs. ~\cite{Dixon:1993xd, Dixon:2004za}. These references found that the $G^3$ and $G^2$ amplitudes do not interfere with each other unless there are at least 3 jets in the final states.  Our amplitudes for $O_3$ must reproduce these amplitudes in the limit of zero Higgs momentum, resulting in vanishing $O_1$-$O_3$ interference. The above references also proposed MHV formulas for $n$-gluon $G^3$ amplitudes involving 3 minus-helicities and $n-3$ plus helicities. We will verify that these MHV formulas hold for the $O_3$ $gggh$ and $ggggh$ amplitudes, i.e. $G^3$ amplitudes at non-zero (and non-lightlike) momentum insertion. This is  expected, as Ref. ~\cite{Dixon:1993xd, Dixon:2004za} already found MHV formulas for the $G^2$ operator to be valid at finite momentum, for Higgs production in the $m_t \to \infty$ limit. 

For convenience, we will first give the lowest-order $gggh$ amplitude for Higgs plus jet production again, in helicity amplitude notation rather than tensor structure notation. The $O_1$ contributions, proportional to $C_1$, are
\begin{align}
im^{O_1}\left(1^{+},2^{+},3^{+}, h\right) & = \frac{2g_{s}m_{h}^{4}}{\langle12\rangle\langle23\rangle\langle31\rangle}, \label{o1+++} \\
im^{O_1}\left(1^{-},2^{+},3^{+}, h\right) &= -\frac{2 g_s [23]^{4}}{[12][23][31]} \label{o1-++}.
\end{align}
The $O_3$ contributions, proportional to $C_3$, are
\begin{align}
im^{O_3}\left(1^{+},2^{+},3^{+}, h\right) & = \frac{-3[12][23][31]}{\Lambda^2}, \label{o3+++} \\
im^{O_3}\left(1^{-},2^{+},3^{+}, h\right) & =0 \label{o3-++},
\end{align}
in agreement with Ref. ~\cite{Neill:2009mz}.
As $p_T$ becomes large, in the Higgs rest frame, the initial and final state jets become much more energetic than the Higgs, so the $m_h\rightarrow 0$
 limit of the above amplitudes, Eqs. \eqref{o1+++}-\eqref{o3-++}, is particularly interesting. In this limit, the $-++$ amplitude is non-zero for $O_1$, but vanishes for $O_3$, so there is no interference between $O_1$ and $O_3$ for this helicity configuration. Meanwhile, the $+++$ amplitude is non-zero as $m_h\rightarrow 0$  for $O_3$, but vanishes as a quartic power in the $m_h\rightarrow 0$
  limit for $O_1$, as seen in Eq. \eqref{o1+++}.  Therefore, we expect the $gggh$ amplitude to not receive large enhancements from the dimension-7 $O_3$ operator at large $p_T$, which means the $m_t \to \infty$ approximation should work well for Higgs differential distribution even at moderately large $p_T$.

Now we will give the $ggggh$ tree amplitudes for $O_3$. They are:
\begin{align}
im^{O_3} \left(1^+, 2^+, 3^+, 4^+, h \right) &= \frac{g_s} {\langle 12 \rangle \langle 23 \rangle \langle 34 \rangle \langle 41 \rangle}
\left( 3i \, S_{12} S_{23} S_{34} - \frac 3 2 i \, S_{12} [31] \langle 1 \slashed{p_H} 2 ] \langle 23 \rangle - \nonumber \right. \\
& \left. \quad \frac 3 2 i S_{12} [42] \langle 2 \slashed{p_H} 1] \langle 14 \rangle \right) + 3\ \rm{\ cyclic\ permutations\ of} \nonumber \\
& \quad (1\to 2\to 3\to 4\to 1), \label{o3++++}\\
im^{O_3} \left( 1^-, 2^-, 3^-, 4^+, h \right) &= \frac {3i\, g_s \langle 12 \rangle^2 \langle 23 \rangle^2 \langle 34 \rangle^2 } {\langle 12 \rangle \langle 23 \rangle \langle 34 \rangle \langle 41 \rangle}, \label{o3---+} \\
im^{O_3} \left( 1^-, 2^-, 3^+, 4^+, h \right) &= 0 \label{o3--++}.
\end{align}

We comment on the massless Higgs limit again. For the $--++$ helicity configuration, the $O_3$ contribution vanishes, while for the $++++$ helicity configuration, the $O_1$ contribution \cite{Dawson:1991au,Kauffman:1996ix} vanishes like a quartic power in the massless Higgs limit. However, for the $---+$ helicity configuration, neither the $O_3$ nor $O_1$ contribution vanishes in the limit $m_h \to 0$ (though the latter vanishes in the limit $p_h \to 0$), so the $O_1$-$O_3$ non-interference at high $p_T$ is no longer true at NLO. 

The amplitudes in Eqs. \eqref{o3+++} and \eqref{o3---+} are unchanged from the MHV formulas for $G^3$ at zero momentum in Ref. ~\cite{Dixon:1993xd,Dixon:2004za}. Furthermore, Refs. \cite{Dixon:2004za, Broedel:2012rc} explored the use of CSW rules \cite{Cachazo:2004kj} to build non-MHV amplitudes from MHV sub-amplitudes for the $G^3$ operator. We confirm that the $++++$ amplitude in Eq. \eqref{o3++++} agrees with the CSW construction with $G^3$ inserted at non-zero momentum. The vanishing of the $--++$ amplitude in Eq. \eqref{o3--++} is explained by the fact that this helicity configuration cannot be built from MHV sub-amplitudes \cite{Dixon:2004za, Broedel:2012rc}.

We have checked that the squared matrix elements from the helicity amplitudes, presented in this section and Appendix \ref{sec:appb}, agree with the automated tree-level calculation by MadGraph5\_aMC@NLO \cite{Alwall:2014hca}, using a UFO model file \cite{Degrande:2011ua} for the dimension-7 operators which we created using FeynRules \cite{Alloul:2013bka}.

\label{nloreal}
\clearpage
\section{Phenomenology}
\label{lhcpheno} 
In this section, we present LO, ${\cal O}(\alpha_s^3)$,  and NLO, 
${\cal O}(\alpha_s^4) $, results  for the Higgs transverse momentum distributions 
resulting from the effective operators,
using the basis of Eq. \ref{ldef1}.  All curves use NLO CJ12 PDFs \cite{Owens:2012bv} with $\mu_F=\mu_R=m_h=126$ GeV, 
$m_t=173$ GeV, and
the 2-loop evolution of $\alpha_s$, with $\alpha_s (126\ {\rm GeV})=0.112497$.
The $O_1$ contribution, with $C_1$ defined in Eq. \eqref{c1pole} to include $\mathcal O(m_h^2 / m_t^2)$ corrections, is equivalent to the $m_t \to \infty$ result rescaled by an overall correction factor. The sum of all contributions, from $O_1$, $O_3$, $O_5$, and the gluon self-interaction operators in Section \ref{sec:self}, gives the full result up to $\mathcal O(m_h^4 / m_t^4)$ corrections in the SM limit.  We use the SM values for the $C_i$ in our plots,
but the 
individual results can be trivially rescaled for BSM coefficients.

\subsection{LO results}
\begin{figure}
\begin{centering}
\includegraphics[scale=0.38]{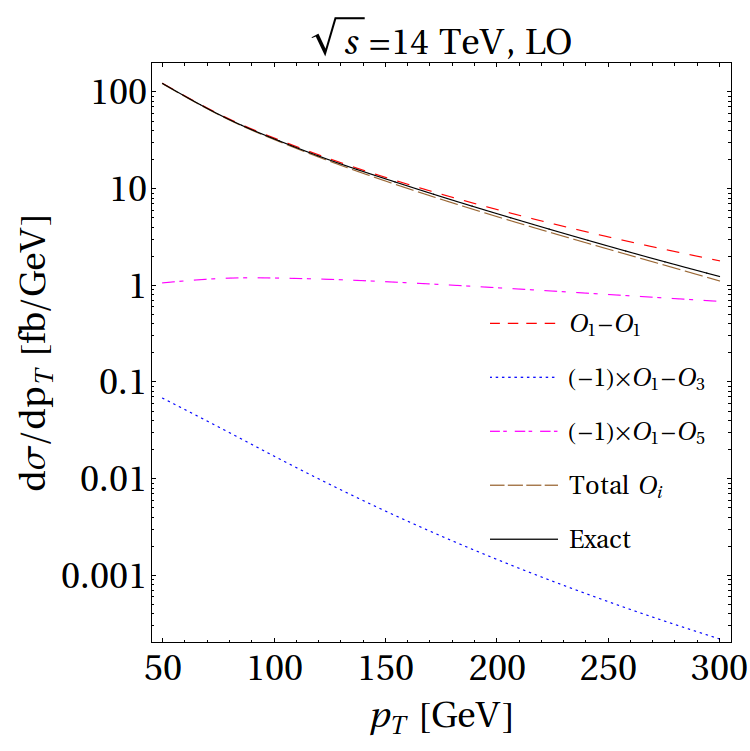}
\par\end{centering}
\caption{Leading order Higgs transverse momentum distributions from the dimension-5 and
dimension-7 EFT operators  for Higgs plus jet production at LO using CJ12 NLO
PDFs with $\mu_R=\mu_F=m_h$.  The curves use the 
${\cal O}(\alpha_s)$
SM values of the $C_i$ and include terms to ${\cal O}\left( 1/m_t^2 \right)$. }
\label{fig:LO}
\end{figure}
\begin{figure}
\begin{centering}
\includegraphics[scale=0.43]{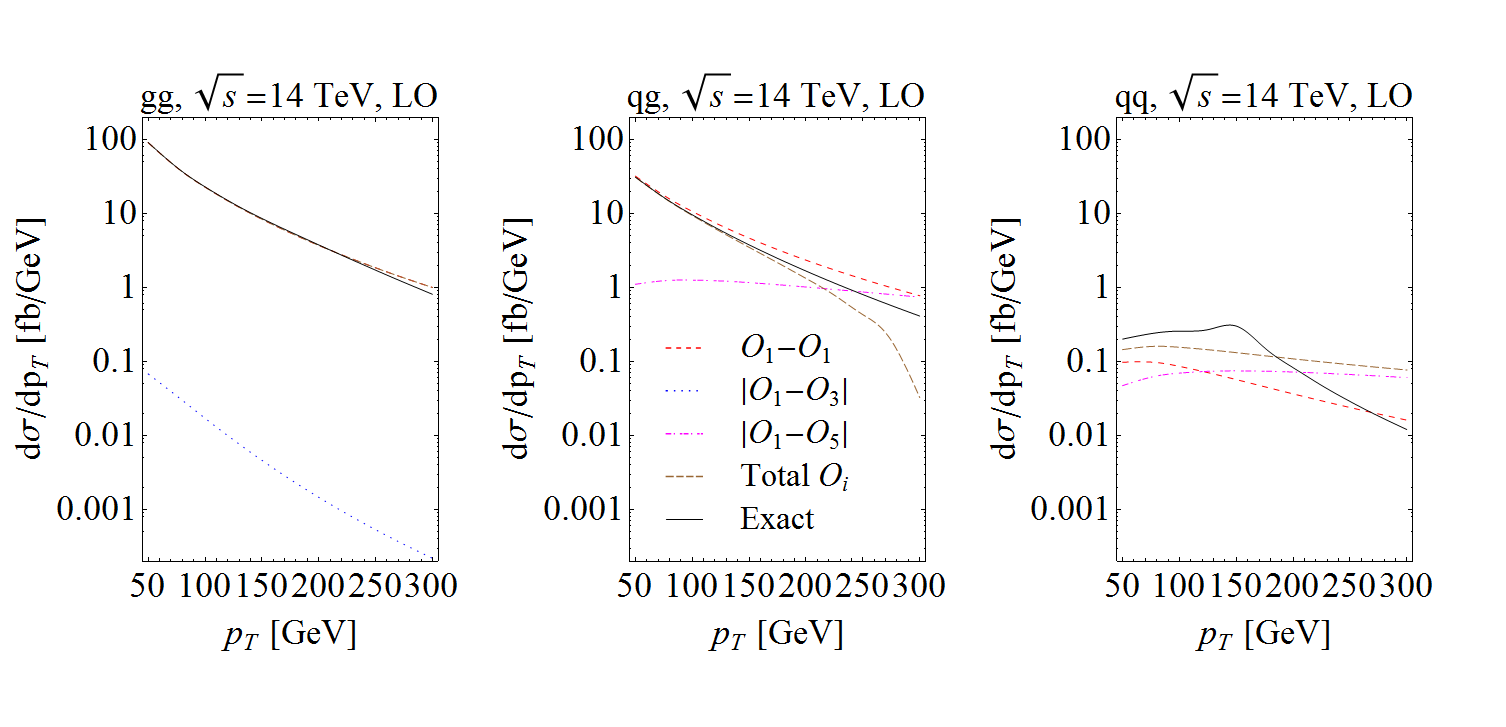}
\par\end{centering}
\caption{Leading order Higgs transverse momentum distributions from the dimension-5 and
dimension-7 EFT operators for Higgs plus jet production at LO using CJ12 NLO
PDFs with $\mu_R=\mu_F=m_h$.  The curves use the 
${\cal O}(\alpha_s)$
SM values of the $C_i$ and include terms to ${\cal O}\left( 1/m_t^2 \right)$. Contributions from $gg$, $qg$, and $qq$ partonic channels are shown separately. }
\label{fig:LO-by-channel}
\end{figure}
At LO, $O_3$ does not contribute to quark channels and $O_5$ does not contribute to the $gg$ channel.
In Fig. \ref{fig:LO}, we plot the LO $p_T$ distribution 
resulting
from the individual operators, and in Fig. \ref{fig:LO-by-channel}, the same plot is broken up into different partonic channels.
The curves labeled as $O_i$-$O_j$ are proportional to $C_i C_j$, where in this section we use the 
${\cal O}(\alpha_s)$ results for the SM $C_i^{\rm SM,pole}$.  
We can see that the $O_1$-$O_1$ result declines as $p_T$ increases
 due to the decrease of the  $gg$ parton luminosity function, 
 while the $O_1$-$O_5$ interference term (which is negative) grows in relative 
 significance at large $p_T$ due to the effects of terms of ${\cal O}(p_T^2 / m_t^2)$
 in the quark-gluon channel.
 The $O_1$-$O_3$ interference term declines even more rapidly than the $O_1$ result
 at high $p_T$, due to the non-interference of the tree-level amplitudes from $O_1$ and $O_3$ in the soft Higgs limit.
 As seen in the real emission section, at tree-level the two operators cannot interfere in the soft Higgs limit unless there are 3 or more jets in the final state.  Also shown is the exact LO result of Ref. \cite{Ellis:1987xu}, 
 including the effects of the top loop exactly.  As made clear also in Ref. \cite{Harlander:2013oja}, 
 the exact and the EFT results diverge for $p_T>150~$GeV.\footnote{After accounting for differing
 input parameters and basis for the dimension-7 operators, our results are in agreement
 with Ref.  \cite{deFlorian:2011xf}. }

Since for LO diagrams without external external quark lines, $O_3$ is the only needed operator that is not from a rescaling of the $m_t \to \infty$ limit, we have an explanation for the excellent agreement between the $O_1$ result and the exact result in the $gg$ channel shown in Fig. \ref{fig:LO}, even at rather large $p_T$.  For the qg-channel, on the other hand, the growing importance of $O_5$ explains the much worse agreement between the EFT result and the exact result at large $p_T$. At small $p_T$, though, the tree-level $qg\to qh$ amplitude factorizes into the collinear splitting $q\to qg$ and the on-shell $gg\to h$ amplitude, which explains the good agreement between the $O_1$ result and the exact result in the qg-channel.  For the $qq$ channel which neither enjoys the special properties  of the $O_3$ helicity amplitudes nor factorizes into gluon sub-amplitudes, we see that the $m_t\to \infty$ approximation with scaling breaks down even at low $p_T \sim 50$ GeV.
\begin{figure}
\begin{centering}
\includegraphics[scale=0.2]{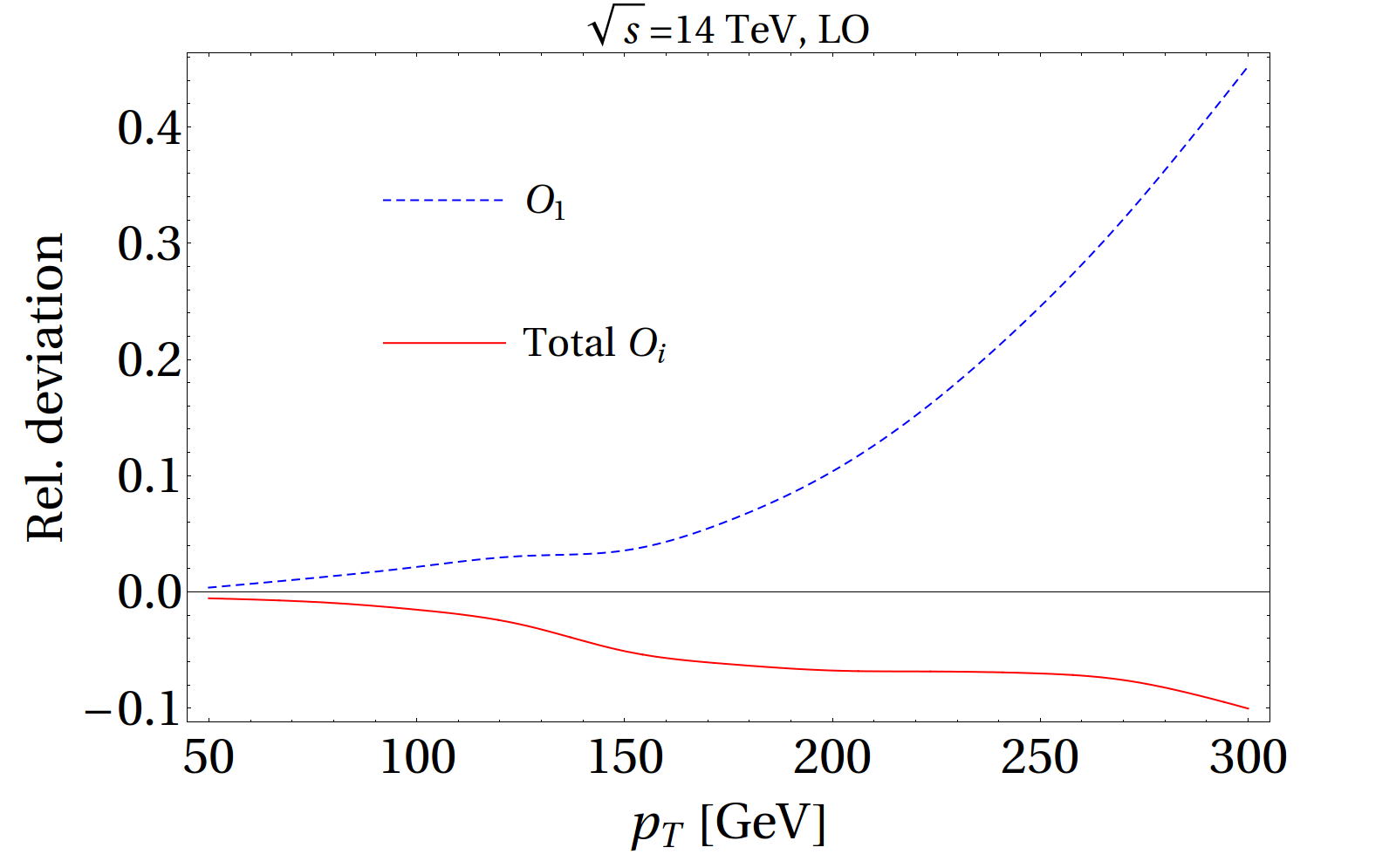}
\par\end{centering}
\caption{Deviations of the EFT predictions 
including all dimension-5 and dimension-7 operators (solid curve)
from the exact result for Higgs plus jet production at LO using CJ12 NLO
PDFs with $\mu_R=\mu_F=m_h$.  The curves use the 
${\cal O}(\alpha_s)$
SM values of the $C_i$ and include terms to ${\cal O}\left( 1/m_t^2 \right)$.  The dotted curve includes only the contribution from $O_1$.}
\label{fig:LO-deviations}
\end{figure}
\begin{figure}
\begin{centering}
\includegraphics[scale=0.4]{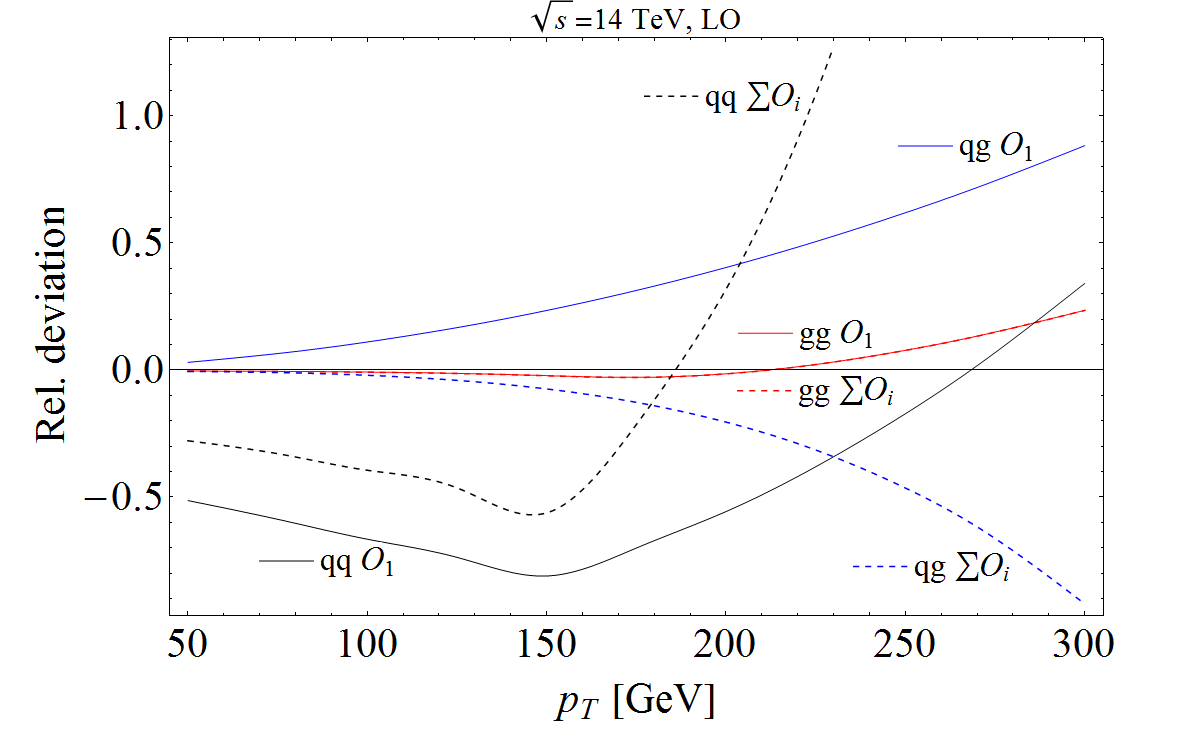}
\par\end{centering}
\caption{Deviations of the EFT predictions from the exact results (dotted curves) , broken up into partonic channels,
for Higgs plus jet production at LO using CJ12 NLO
PDFs with $\mu_R=\mu_F=m_h$.  The curves use the ${\cal O}(\alpha_s)$ SM values of the $C_i$ and include terms to ${\cal O}\left( 1/m_t^2 \right)$.  The solid  curves includes only the contribution from $O_1$.
The red dashed and red solid curves are indistinguishable.}
\label{fig:LO-deviations-by-channel}
\end{figure}
In Figs. \ref{fig:LO-deviations} and \ref{fig:LO-deviations-by-channel} we plot the deviation of the $O_1$ result and the total result from the exact result. We again see the remarkably tame deviation in the gg channel
from the exact result,  while observing that including all dimension-7 operators gives a better approximation to the exact $p_T$ distribution than including the effects of $O_1$ alone, especially for $p_T < m_h$.
\clearpage
\subsection{Numerical accuracy at NLO}
\begin{figure}
\begin{centering}
\includegraphics[scale=0.6]{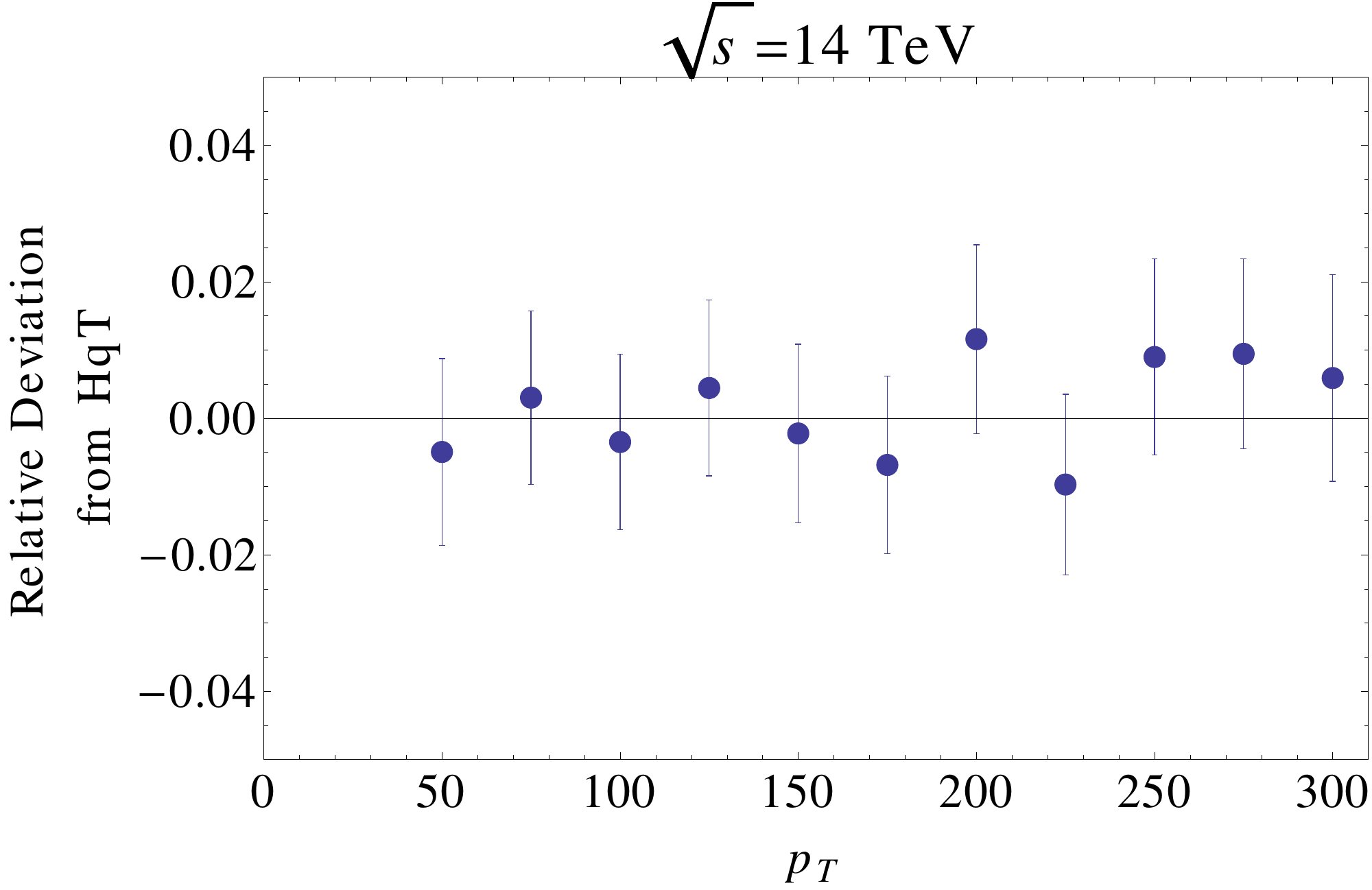}
\par\end{centering}
\caption{Deviation of our  NLO result for the  Higgs $p_T$ distribution in the large $m_t$ limit from the results of the HqT 2.0
 program \cite{deFlorian:2011xf} using $\delta_s=10^{-3}$, and
$\delta_c = \delta_s / 200$ for $p_T \geq 75\ {\rm GeV}$ and $\delta_c = \delta_s / 400$ for $p_T = 50\ {\rm GeV}$.}
\label{fig:dev-from-hqt}
\end{figure}
\begin{figure}
\begin{centering}
\includegraphics[scale=0.25]{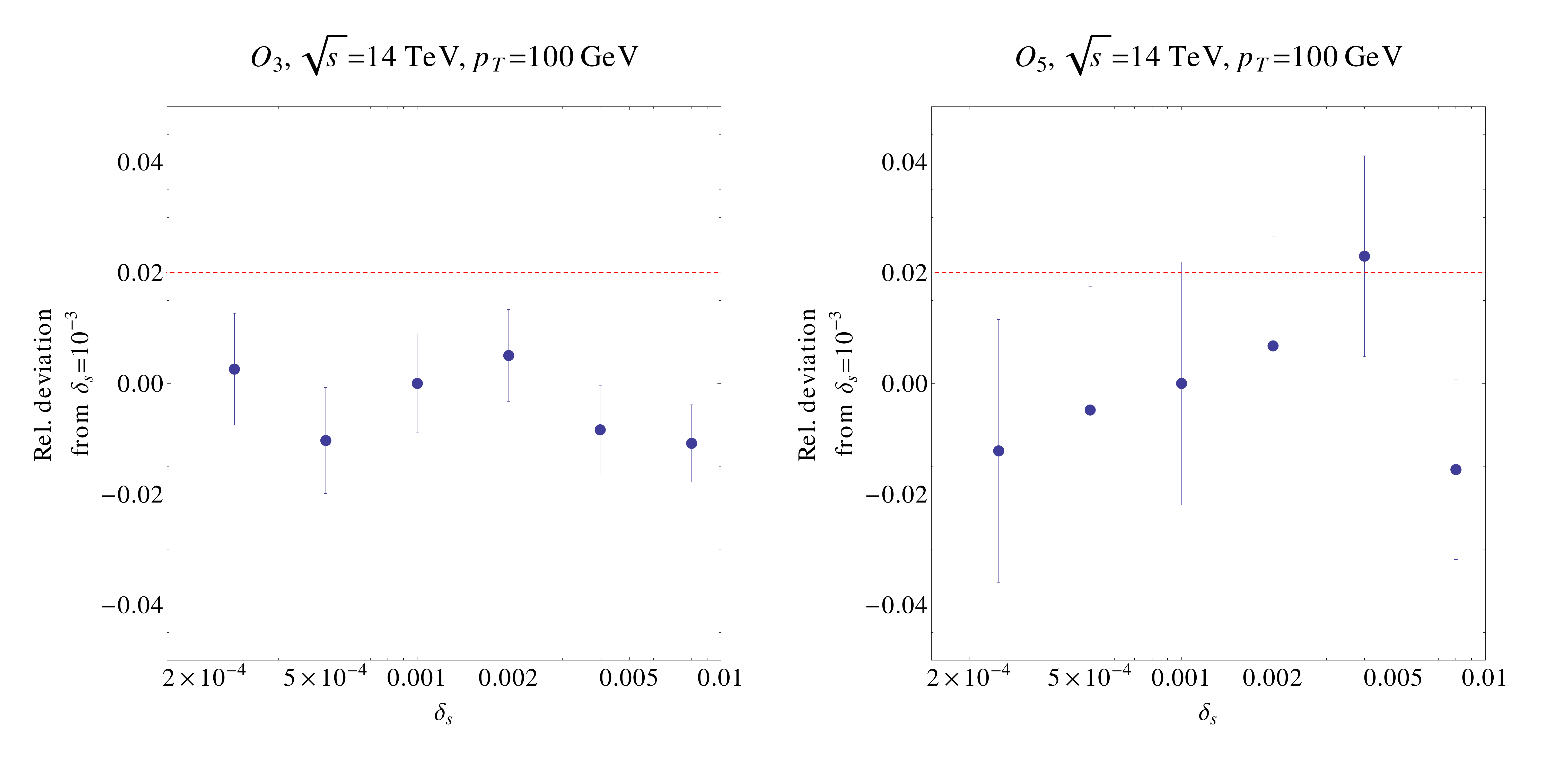}
\par\end{centering}
\caption{Dependence of the NLO result for the Higgs $p_T$ distribution on the soft cutoff, $\delta_s$,
including only the interference of $O_1$ with $O_3$ (LHS) and $O_1$ with $O_5$ (RHS).
The collinear cutoff is taken to be $\delta_c=5\times 10^{-6}$. The result with $\delta_s = 10^{-3}$ is normalized to 1.}
\label{fig:o3o5-deltas}
\end{figure}
Our NLO results are derived using phase space slicing with 2 cut-offs, $\delta_c$ and $\delta_s$.  
To show the accuracy of our implementation of phase space slicing, in Fig. \ref{fig:dev-from-hqt},  we show the deviation of our NLO result for the $m_t \to \infty$ limit from the result produced by HqT 2.0 \cite{deFlorian:2011xf}. (The
errors are statistical). 
We find agreement at the percent level. 
The variation of $d\sigma / d p_T$  with $\delta_s$ for the $O_3$ and $O_5$  operators
individually (using the SM ${\cal O}(\alpha_s^2) $ values for the $C_i^{\rm SM,pole}$ coefficients)
is plotted in Fig. \ref{fig:o3o5-deltas} for fixed $\delta_c=5\times 10^{-6}$ and for $p_T=100~$GeV.  We see
that at the percent level, our results are independent of the choice of soft cutoff.  Similarly, we have
verified the  there is no dependence on the collinear cutoff when $\delta_c <<\delta_s$.  Our
results in the following sections use $\delta_c=5 \times 10^{-6}$ (except for the $O_1$ result at $p_T=50.0$ GeV, for which we use one half this value) and $\delta_s=10^{-3}$.  All the plots are made by computing at $\delta p_T = 25$ GeV intervals, joined together by smooth curves, and it should kept in mind that an error of $\sim 1-2\%$ is present.
\subsection{NLO results}
\begin{figure}
\begin{centering}
\includegraphics[scale=0.28]{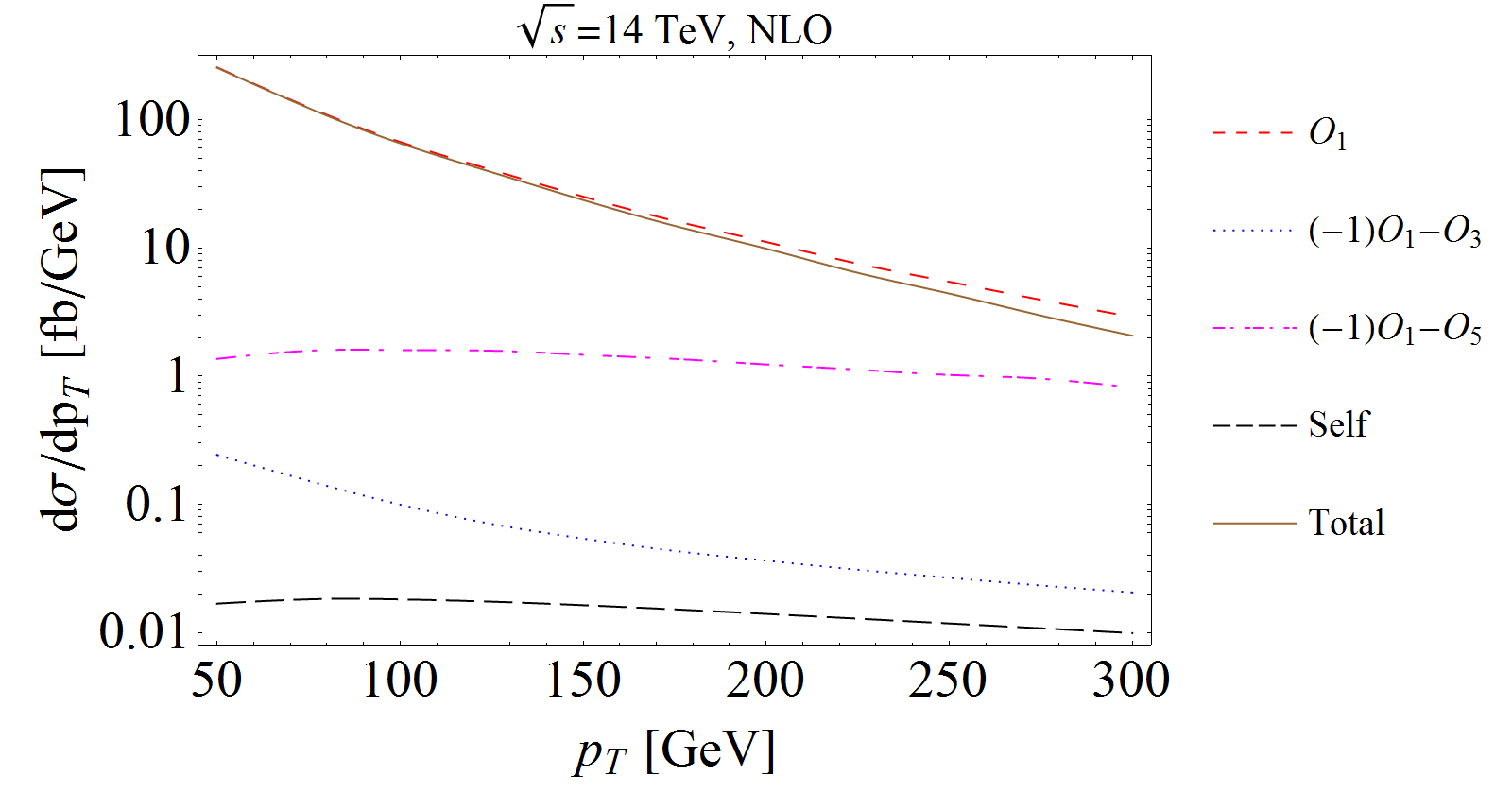}
\par\end{centering}
\caption{Next-to-leading order Higgs transverse momentum distributions from the EFT 
dimension-5 and dimension-7 operators,
using the SM values of $C_i^{\rm SM,pole}$ to ${\cal O}(\alpha_s^2)$ and include terms only to
${\cal O}\left( 1/m_t^2 \right)$.}
\label{fig:NLO}
\end{figure}
In Fig. \ref{fig:NLO},  we plot the contributions of the dimension-5 and dimension-7
 EFT operators to the NLO $p_T$ distributions.  The NLO
plots use the ${\cal O}(\alpha_s^2)$ expressions for the $C_i^{\rm SM,pole}$ and include terms only to
${\cal O}\left( 1/m_t^2 \right)$. 
Compared with the LO plot in Fig. \ref{fig:LO}, an important change is that the dimension-7 $O_3$ contribution no longer shows the property of declining faster than the dimension-5 $O_1$ contribution (because interference between $O_3$ and $O_1$ amplitudes in the soft Higgs limit starts at NLO), although $O_5$ is still dominant at large $p_T$.   The curve labeled ``self'' is the small contribution from the $O(1/m_t^2)$ gluon self-couplings
of Eq. \ref{Lself}.
\begin{figure}
\begin{centering}
\includegraphics[scale=0.3]{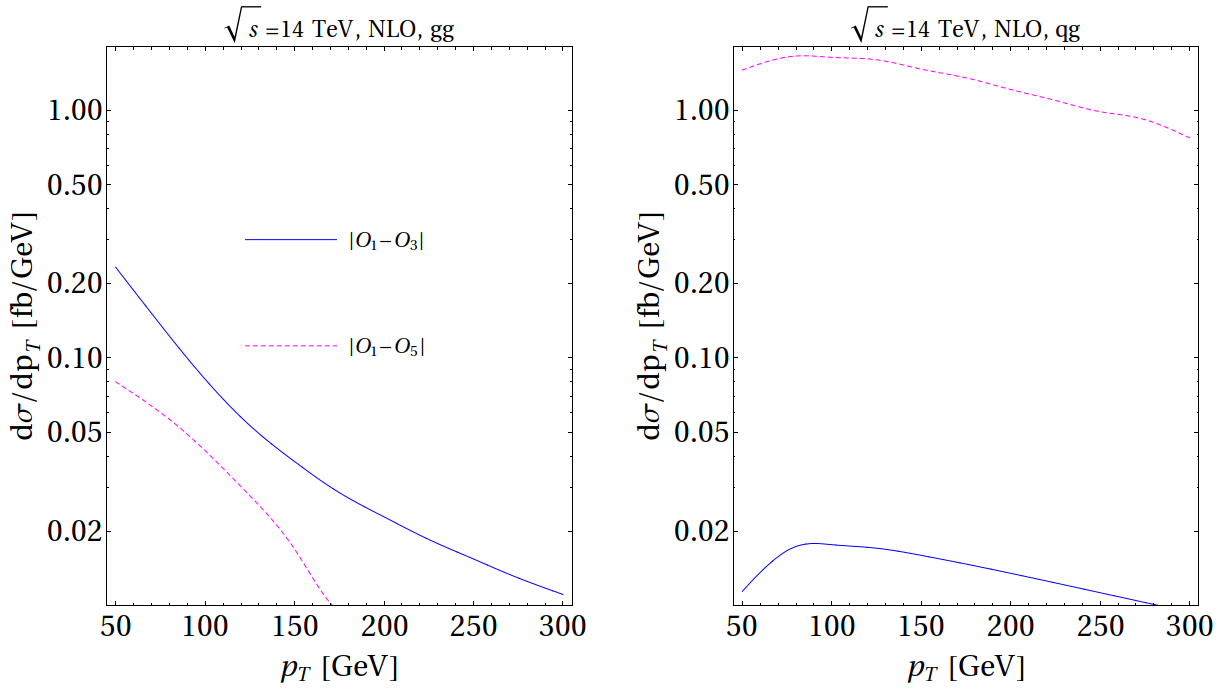}
\par\end{centering}
\caption{Comparison of the sizes of $O_3$ and $O_5$ contributions in the $gg$ and $qg$ channels at NLO.} 
\label{fig:NLO-by-channel}
\end{figure}
The dimension-7 contributions to the $gg$ and $qg$ individual channels are shown in Fig. \ref{fig:NLO-by-channel}. In the $gg$ channel, the $O_5$ operator starts to have non-vanishing contribution at NLO, but the contribution remains small compared with $O_3$,  partly because $O_5$ only affects diagrams involving external quark legs or internal quark loops. In the $qg$ channel, the $O_3$ operator starts to have non-vanishing contribution at NLO, but the contribution remains small compared with $O_5$. Therefore, we should still associate $O_3$ primarily with the $gg$ channel, and $O_5$ primarily with channels involving initial-state quarks.

\begin{figure}
\begin{centering}
\includegraphics[scale=0.3]{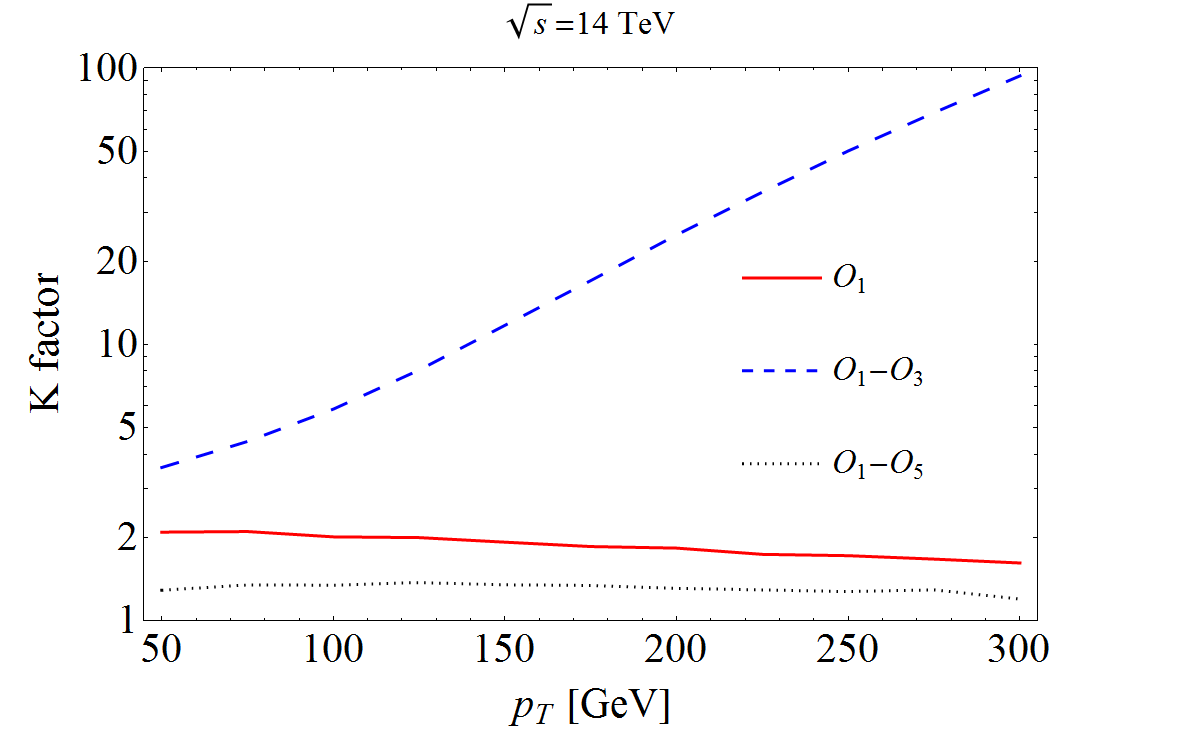}
\par\end{centering}
\caption{The NLO $p_T$-dependent K-factor for each of the operators, as defined in
Eq. \ref{kdef}.}
\label{fig:K-factor-by-operator}
\end{figure}
In order to quantify the size of our results, we define a $p_T$ dependent K-factor:
\begin{equation}
K(p_T)={ {d\sigma\over dp_T}(\rm{NLO})\over {d\sigma\over dp_T} (\rm{LO})}\, ,
\label{kdef}
\end{equation}
where in our plots both the NLO and LO curves use CJ12 PDFs with the 2-loop evolution of $\alpha_s$.
We plot the K factor separately for the contributions from $O_1$ and for the contributions from the
interference of $O_1$ with $O_3$ and $O_5$.   The results use the SM values of $C_i^{\rm SM,pole}$, but
can be rescaled appropriately for BSM models. 
In Fig. \ref{fig:K-factor-by-operator},   we  see that the NLO K-factors for $O_1$ and $O_5$ are always of order unity, while the $O_3$ K-factor reaches huge values at large $p_T$, reflecting the fact that the vanishing interference between the $O_1$ and $O_3$ helicity amplitudes in the soft Higgs limit no longer holds at one-loop level.

\begin{figure}
\begin{centering}
\includegraphics[scale=0.3]{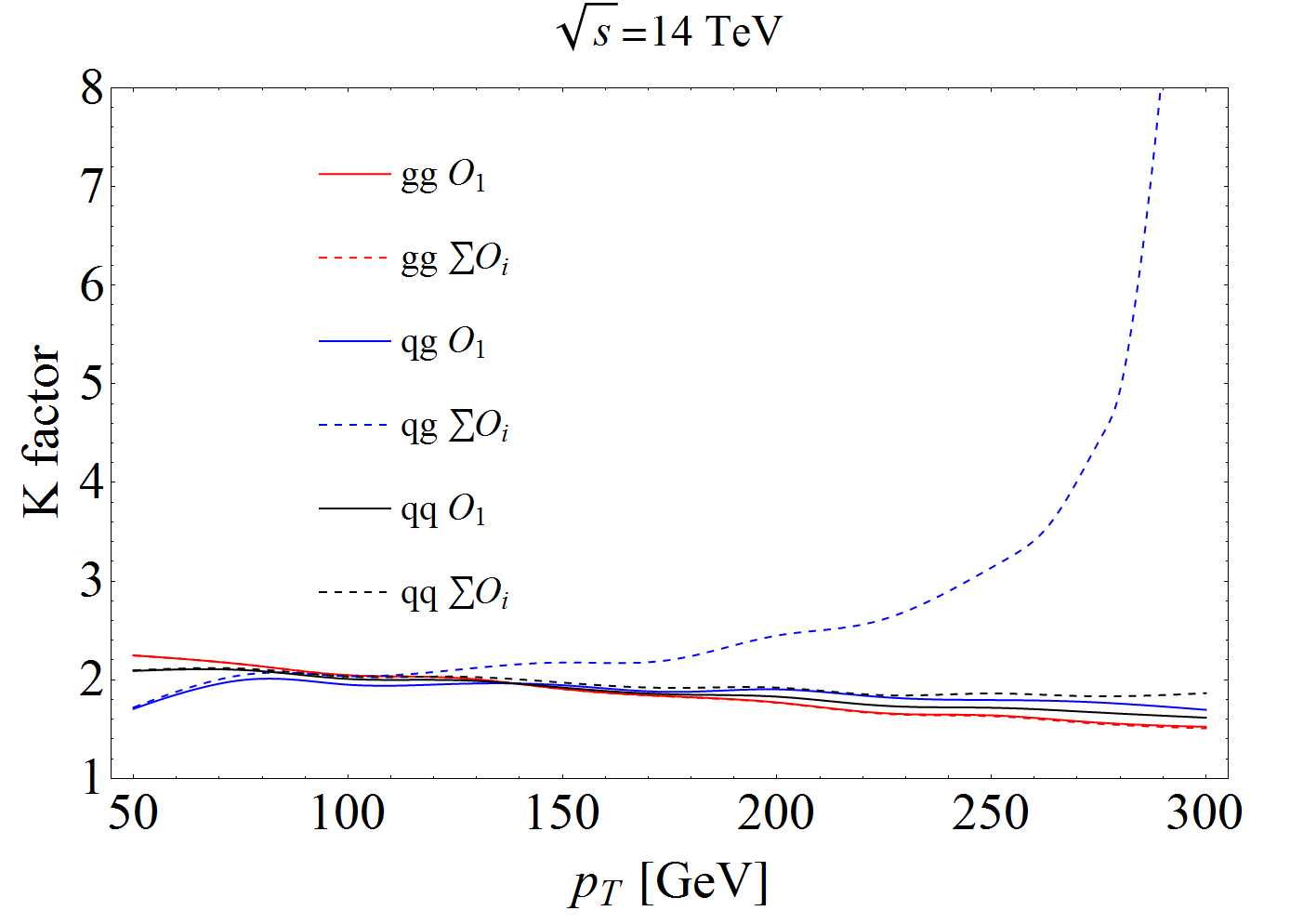}
\par\end{centering}
\caption{The NLO $p_T$-dependent K-factor, broken up into partonic channels.}
\label{fig:K-factor-by-channel}
\end{figure}
\begin{figure}
\begin{centering}
\includegraphics[scale=0.25]{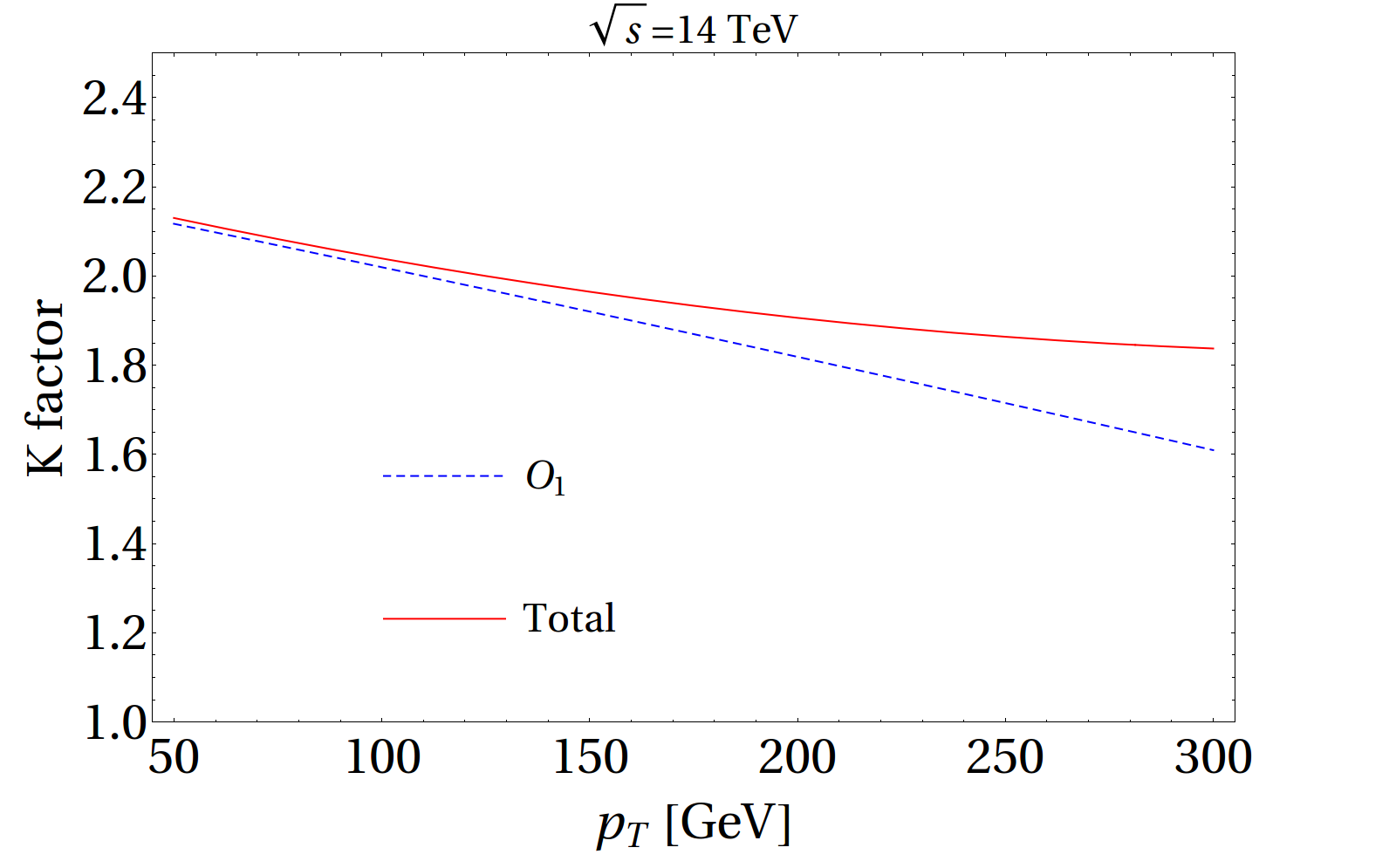}
\par\end{centering}
\caption{The NLO $p_T$-dependent K-factor, broken up into partonic channels, summed over all partonic channels.}
\label{fig:K-factor}
\end{figure}
In Fig. \ref{fig:K-factor-by-channel},  we show the NLO 
$p_T$ dependent
K-factors for each partonic channel. We can see that in going from 
the contribution of only $O_1$ to the sum of the contributions from all operators, the K-factor hardly
 changes in the $gg$-channel, while there are significant changes in the $qg$ and 
 $qq$  channels. This is not surprising given the high $p_T$ suppression of the  $O_3$  contribution and the lack of an $O_5$ contribution in the all-gluon channel at LO, while the NLO effects are not large enough to destroy the agreement with 
 the contribution of $O_1$ alone. In Fig. \ref{fig:K-factor} we observe that when all partonic channels are summed up, the K-factor only shows modest changes \cite{Grazzini:2013mca,Harlander:2012hf}  due to the dominance of the $gg$ channel.

Our K-factors plots are for SM Higgs production, with the non-logarithmic terms $\hat C^{(1)}_3$ and $\hat C^{(1)}_5$ in Eq. \eqref{c3hat}, \eqref{c5hat} set to zero. It is straightforward to scale the K factors to reflect the effects of BSM physics. Define the K-factors corresponding to $O_i$ as $K^i$, and define the expansion in $\alpha_s$ for SM and BSM coefficients,
\begin{align}
C_i^{\rm SM} &= \alpha_s C_i^{(0,{\rm SM})} + \alpha_s^2 C_i^{(1,{\rm SM})}, \nonumber \\
C_i^{\rm BSM} &= \alpha_s C_i^{(0,{\rm BSM})} + \alpha_s^2 C_i^{(1,{\rm BSM})}.
\end{align}
The K-factor for a BSM model can be derived to $O(\alpha_s)$ by the rescaling,
\begin{align}
\frac{K^{1,{\rm BSM}}} {K^{1,{\rm SM}}} &= 1 + 2 \alpha_s \left( \frac {C_1^{(1,{\rm BSM})}} {C_1^{(0,{\rm BSM})}} - \frac {C_1^{(1,{\rm SM})}} {C_1^{(0,{\rm SM})}} \right), \nonumber \\
\frac{K^{1,{\rm BSM}}} {K^{1,{\rm SM}}} &= 1 + \alpha_s \left( \frac {C_1^{(1,{\rm BSM})}} {C_1^{(0,{\rm BSM})}} - \frac {C_1^{(1,{\rm SM})}} {C_1^{(0,{\rm SM})}} + \frac {C_5^{(1,{\rm BSM})}} {C_5^{(0,{\rm BSM})}} - \frac {C_5^{(1,{\rm SM})}} {C_5^{(0,{\rm SM})}} \right), \nonumber \\
\frac{K^{1,{\rm BSM}}} {K^{1,{\rm SM}}} &= 1 + \alpha_s \left( \frac {C_1^{(1,{\rm BSM})}} {C_1^{(0,{\rm BSM})}} - \frac {C_1^{(1,{\rm SM})}} {C_1^{(0,{\rm SM})}} + \frac {C_3^{(1,{\rm BSM})}} {C_3^{(0,{\rm BSM})}} - \frac {C_3^{(1,{\rm SM})}} {C_3^{(0,{\rm SM})}} \right)\,.
\end{align}

\clearpage
\section{Conclusion}
\label{concs}

We used an effective field theory containing strong gluon-Higgs-quark operators to dimension-7
 to parameterize either non-SM couplings or the effect of a finite top mass within the SM.
We calculated the NLO, ${\cal O}(\alpha_s^4)$, contribution to the $p_T$ spectrum for Higgs plus jet production,
including effects of ${\cal O}(1/\Lambda^2)$, for arbitrary values of the coefficients, $C_i$, of the effective Lagrangian.
There are $3$ dimension-7 operators which contribute to Higgs plus jet production:
$O_6 \cong m_h^2 O_1$, $O_3$, and $O_5$.  The operator $O_6$ rescales the overall gluon fusion rate for Higgs production and is constrained to be close to the SM value.
The contribution from $O_3$, mainly in the $gg$ channel, is suppressed at LO for large $p_T$ since it vanishes in the soft Higgs limit, and remains numerically small at NLO, making it difficult to observe new physics in this channel, and also suppressing the dependence on the top quark mass.  The contribution from $O_5$, which is mainly in the $qg$ channel, is significant at large $p_T$.  Hence,  BSM
physics will be most readily accessible if it contains a significant enhancement of $C_5$ over the SM value. We studied the renormalization of the dimension-7 operators, which makes it possible to regulate the UV divergence of the one-loop amplitudes and to use renormalization group running, from the BSM scale down to the Higgs mass scale, to resum large logarithms.

 When the operator coefficients are set to their SM values, we obtain the
 ${\cal}O(1/m_t^2)$ corrections to the NLO rate for Higgs plus jet production, modulo the non-logarithmic terms in the NLO matching coefficients in Eqs. \eqref{c3hat},\eqref{c5hat} to be presented shortly in a forthcoming work. These corrections are well
 behaved in the $gg$ channel, but become increasingly large in the $qg$ channel as $p_T$ is increased
 above $m_h$.  This observation is in agreement with Ref. \cite{Harlander:2012hf}.
   We present $p_T$ dependent $K$ factors which 
 can be easily rescaled to include  BSM physics.  

\begin{acknowledgments}
The work of SD and IL is supported  the U.S. Department of Energy under grant
No.~DE-AC02-98CH10886.  The work of MZ is supported by NSF grant PHY-1316617.
We thank Lance Dixon, Duff Neill, and George Sterman for useful discussions. 
\end{acknowledgments}
\newpage
\bibliographystyle{unsrt}
\bibliography{hjet}

\begin{thebibliography}{10}

\bibitem{ATLAS-CONF-2014-009}
{Updated coupling measurements of the Higgs boson with the ATLAS detector using
  up to 25 fb$^{-1}$ of proton-proton collision data}.
\newblock Technical Report ATLAS-CONF-2014-009, CERN, Geneva, Mar 2014.

\bibitem{CMS-PAS-HIG-14-009}
{Precise determination of the mass of the Higgs boson and studies of the
  compatibility of its couplings with the standard model}.
\newblock Technical Report CMS-PAS-HIG-14-009, CERN, Geneva, 2014.

\bibitem{Dittmaier:2011ti}
S.~Dittmaier et~al.
\newblock {Handbook of LHC Higgs Cross Sections: 1. Inclusive Observables}.
\newblock 2011.

\bibitem{Dittmaier:2012vm}
S.~Dittmaier, S.~Dittmaier, C.~Mariotti, G.~Passarino, R.~Tanaka, et~al.
\newblock {Handbook of LHC Higgs Cross Sections: 2. Differential
  Distributions}.
\newblock 2012.

\bibitem{Kniehl:1995tn}
Bernd~A. Kniehl and Michael Spira.
\newblock {Low-energy theorems in Higgs physics}.
\newblock {\em Z.Phys.}, C69:77--88, 1995.

\bibitem{Spira:1995rr}
M.~Spira, A.~Djouadi, D.~Graudenz, and P.M. Zerwas.
\newblock {Higgs boson production at the LHC}.
\newblock {\em Nucl.Phys.}, B453:17--82, 1995.

\bibitem{Dawson:1990zj}
S.~Dawson.
\newblock {Radiative corrections to Higgs boson production}.
\newblock {\em Nucl.Phys.}, B359:283--300, 1991.

\bibitem{ArkaniHamed:2002qy}
N.~Arkani-Hamed, A.G. Cohen, E.~Katz, and A.E. Nelson.
\newblock {The Littlest Higgs}.
\newblock {\em JHEP}, 0207:034, 2002.

\bibitem{Contino:2003ve}
Roberto Contino, Yasunori Nomura, and Alex Pomarol.
\newblock {Higgs as a holographic pseudoGoldstone boson}.
\newblock {\em Nucl.Phys.}, B671:148--174, 2003.

\bibitem{Low:2010mr}
Ian Low and Alessandro Vichi.
\newblock {On the production of a composite Higgs boson}.
\newblock {\em Phys.Rev.}, D84:045019, 2011.

\bibitem{Carena:2013qia}
M.~Carena, S.~Heinemeyer, O.~StŒl, C.E.M. Wagner, and G.~Weiglein.
\newblock {MSSM Higgs Boson Searches at the LHC: Benchmark Scenarios after the
  Discovery of a Higgs-like Particle}.
\newblock {\em Eur.Phys.J.}, C73:2552, 2013.

\bibitem{Carena:2013iba}
Marcela Carena, Stefania Gori, Nausheen~R. Shah, Carlos~E.M. Wagner, and
  Lian-Tao Wang.
\newblock {Light Stops, Light Staus and the 125 GeV Higgs}.
\newblock {\em JHEP}, 1308:087, 2013.

\bibitem{Grojean:2013nya}
Christophe Grojean, Ennio Salvioni, Matthias Schlaffer, and Andreas Weiler.
\newblock {Very boosted Higgs in gluon fusion}.
\newblock {\em JHEP}, 1405:022, 2014.

\bibitem{Azatov:2013xha}
Aleksandr Azatov and Ayan Paul.
\newblock {Probing Higgs couplings with high $p_T$ Higgs production}.
\newblock {\em JHEP}, 1401:014, 2014.

\bibitem{Low:2009di}
Ian Low, Riccardo Rattazzi, and Alessandro Vichi.
\newblock {Theoretical Constraints on the Higgs Effective Couplings}.
\newblock {\em JHEP}, 1004:126, 2010.

\bibitem{Buschmann:2014twa}
Malte Buschmann, Christoph Englert, Dorival Goncalves, Tilman Plehn, and
  Michael Spannowsky.
\newblock {Resolving the Higgs-Gluon Coupling with Jets}.
\newblock {\em Phys.Rev.}, D90:013010, 2014.

\bibitem{Schlaffer:2014osa}
Matthias Schlaffer, Michael Spannowsky, Michihisa Takeuchi, Andreas Weiler, and
  Chris Wymant.
\newblock {Boosted Higgs Shapes}.
\newblock 2014.

\bibitem{Azatov:2011qy}
Aleksandr Azatov and Jamison Galloway.
\newblock {Light Custodians and Higgs Physics in Composite Models}.
\newblock {\em Phys.Rev.}, D85:055013, 2012.

\bibitem{Dawson:2012di}
S.~Dawson and E.~Furlan.
\newblock {A Higgs Conundrum with Vector Fermions}.
\newblock {\em Phys.Rev.}, D86:015021, 2012.

\bibitem{Gillioz:2012se}
M.~Gillioz, R.~Grober, C.~Grojean, M.~Muhlleitner, and E.~Salvioni.
\newblock {Higgs Low-Energy Theorem (and its corrections) in Composite Models}.
\newblock {\em JHEP}, 1210:004, 2012.

\bibitem{Dawson:2012mk}
Sally Dawson, Elisabetta Furlan, and Ian Lewis.
\newblock {Unravelling an extended quark sector through multiple Higgs
  production?}
\newblock {\em Phys.Rev.}, D87:014007, 2013.

\bibitem{Banfi:2013yoa}
Andrea Banfi, Adam Martin, and Veronica Sanz.
\newblock {Probing top-partners in Higgs + jets}.
\newblock 2013.

\bibitem{Neill:2009tn}
Duff Neill.
\newblock {Two-Loop Matching onto Dimension Eight Operators in the Higgs-Glue
  Sector}.
\newblock 2009.

\bibitem{Harlander:2013oja}
Robert~V. Harlander and Tobias Neumann.
\newblock {Probing the nature of the Higgs-gluon coupling}.
\newblock {\em Phys.Rev.}, D88:074015, 2013.

\bibitem{Harlander:2002wh}
Robert~V. Harlander and William~B. Kilgore.
\newblock {Next-to-next-to-leading order Higgs production at hadron colliders}.
\newblock {\em Phys.Rev.Lett.}, 88:201801, 2002.

\bibitem{Ravindran:2003um}
V.~Ravindran, J.~Smith, and W.~L. van Neerven.
\newblock {NNLO corrections to the total cross-section for Higgs boson
  production in hadron hadron collisions}.
\newblock {\em Nucl.Phys.}, B665:325--366, 2003.

\bibitem{Anastasiou:2002yz}
Charalampos Anastasiou and Kirill Melnikov.
\newblock {Higgs boson production at hadron colliders in NNLO QCD}.
\newblock {\em Nucl.Phys.}, B646:220--256, 2002.

\bibitem{Anastasiou:2005qj}
Charalampos Anastasiou, Kirill Melnikov, and Frank Petriello.
\newblock {Fully differential Higgs boson production and the di-photon signal
  through next-to-next-to-leading order}.
\newblock {\em Nucl.Phys.}, B724:197--246, 2005.

\bibitem{Catani:2007vq}
Stefano Catani and Massimiliano Grazzini.
\newblock {An NNLO subtraction formalism in hadron collisions and its
  application to Higgs boson production at the LHC}.
\newblock {\em Phys.Rev.Lett.}, 98:222002, 2007.

\bibitem{Ravindran:2002dc}
V.~Ravindran, J.~Smith, and W.L. Van~Neerven.
\newblock {Next-to-leading order QCD corrections to differential distributions
  of Higgs boson production in hadron hadron collisions}.
\newblock {\em Nucl.Phys.}, B634:247--290, 2002.

\bibitem{Harlander:2009mq}
Robert~V. Harlander and Kemal~J. Ozeren.
\newblock {Finite top mass effects for hadronic Higgs production at
  next-to-next-to-leading order}.
\newblock {\em JHEP}, 0911:088, 2009.

\bibitem{Pak:2009dg}
Alexey Pak, Mikhail Rogal, and Matthias Steinhauser.
\newblock {Finite top quark mass effects in NNLO Higgs boson production at
  LHC}.
\newblock {\em JHEP}, 1002:025, 2010.

\bibitem{Chetyrkin:1997un}
K.G. Chetyrkin, Bernd~A. Kniehl, and M.~Steinhauser.
\newblock {Decoupling relations to O ($\alpha_s^3$) and their connection to
  low-energy theorems}.
\newblock {\em Nucl.Phys.}, B510:61--87, 1998.

\bibitem{Schroder:2005hy}
Y.~Schroder and M.~Steinhauser.
\newblock {Four-loop decoupling relations for the strong coupling}.
\newblock {\em JHEP}, 0601:051, 2006.

\bibitem{Chetyrkin:2005ia}
K.G. Chetyrkin, Johann~H. Kuhn, and Christian Sturm.
\newblock {QCD decoupling at four loops}.
\newblock {\em Nucl.Phys.}, B744:121--135, 2006.

\bibitem{Kramer:1996iq}
Michael Kramer, Eric Laenen, and Michael Spira.
\newblock {Soft gluon radiation in Higgs boson production at the LHC}.
\newblock {\em Nucl.Phys.}, B511:523--549, 1998.

\bibitem{Ellis:1987xu}
R.~Keith Ellis, I.~Hinchliffe, M.~Soldate, and J.J. van~der Bij.
\newblock {Higgs Decay to tau+ tau-: A Possible Signature of Intermediate Mass
  Higgs Bosons at the SSC}.
\newblock {\em Nucl.Phys.}, B297:221, 1988.

\bibitem{Baur:1989cm}
U.~Baur and E.W.~Nigel Glover.
\newblock {Higgs Boson Production at Large Transverse Momentum in Hadronic
  Collisions}.
\newblock {\em Nucl.Phys.}, B339:38--66, 1990.

\bibitem{deFlorian:1999zd}
D.~de~Florian, M.~Grazzini, and Z.~Kunszt.
\newblock {Higgs production with large transverse momentum in hadronic
  collisions at next-to-leading order}.
\newblock {\em Phys.Rev.Lett.}, 82:5209--5212, 1999.

\bibitem{Glosser:2002gm}
Christopher~J. Glosser and Carl~R. Schmidt.
\newblock {Next-to-leading corrections to the Higgs boson transverse momentum
  spectrum in gluon fusion}.
\newblock {\em JHEP}, 0212:016, 2002.

\bibitem{Harlander:2012hf}
Robert~V. Harlander, Tobias Neumann, Kemal~J. Ozeren, and Marius Wiesemann.
\newblock {Top-mass effects in differential Higgs production through gluon
  fusion at order alpha s4}.
\newblock {\em JHEP}, 1208:139, 2012.

\bibitem{Grazzini:2013mca}
Massimiliano Grazzini and Hayk Sargsyan.
\newblock {Heavy-quark mass effects in Higgs boson production at the LHC}.
\newblock {\em JHEP}, 1309:129, 2013.

\bibitem{Bagnaschi:2011tu}
E.~Bagnaschi, G.~Degrassi, P.~Slavich, and A.~Vicini.
\newblock {Higgs production via gluon fusion in the POWHEG approach in the SM
  and in the MSSM}.
\newblock {\em JHEP}, 1202:088, 2012.

\bibitem{Neumann:2014nha}
Tobias Neumann and Marius Wiesemann.
\newblock {Finite top-mass effects in gluon-induced Higgs production with a
  jet-veto at NNLO}.
\newblock 2014.

\bibitem{Keung:2009bs}
Wai-Yee Keung and Frank~J. Petriello.
\newblock {Electroweak and finite quark-mass effects on the Higgs boson
  transverse momentum distribution}.
\newblock {\em Phys.Rev.}, D80:013007, 2009.

\bibitem{Boughezal:2013uia}
Radja Boughezal, Fabrizio Caola, Kirill Melnikov, Frank Petriello, and Markus
  Schulze.
\newblock {Higgs boson production in association with a jet at
  next-to-next-to-leading order in perturbative QCD}.
\newblock {\em JHEP}, 1306:072, 2013.

\bibitem{Becher:2014tsa}
Thomas Becher, Guido Bell, Christian Lorentzen, and Stefanie Marti.
\newblock {The transverse-momentum spectrum of Higgs bosons near threshold at
  NNLO}.
\newblock 2014.

\bibitem{Becher:2013vva}
Thomas Becher, Guido Bell, Christian Lorentzen, and Stefanie Marti.
\newblock {Transverse-momentum spectra of electroweak bosons near threshold at
  NNLO}.
\newblock {\em JHEP}, 1402:004, 2014.

\bibitem{Huang:2014mca}
Fa~Peng Huang, Chong~Sheng Li, Hai~Tao Li, and Jian Wang.
\newblock {Renormalization-group improved predictions for Higgs boson
  production at large $p_T$}.
\newblock 2014.

\bibitem{DelDuca:2001eu}
V.~Del~Duca, W.~Kilgore, C.~Oleari, C.~Schmidt, and D.~Zeppenfeld.
\newblock {Higgs + 2 jets via gluon fusion}.
\newblock {\em Phys.Rev.Lett.}, 87:122001, 2001.

\bibitem{DelDuca:2001fn}
V.~Del~Duca, W.~Kilgore, C.~Oleari, C.~Schmidt, and D.~Zeppenfeld.
\newblock {Gluon fusion contributions to H + 2 jet production}.
\newblock {\em Nucl.Phys.}, B616:367--399, 2001.

\bibitem{Campanario:2013mga}
Francisco Campanario and Michael Kubocz.
\newblock {Higgs boson production in association with three jets via gluon
  fusion at the LHC: Gluonic contributions}.
\newblock {\em Phys.Rev.}, D88(5):054021, 2013.

\bibitem{vanDeurzen:2013rv}
H.~van Deurzen, N.~Greiner, G.~Luisoni, P.~Mastrolia, E.~Mirabella, et~al.
\newblock {NLO QCD corrections to the production of Higgs plus two jets at the
  LHC}.
\newblock {\em Phys.Lett.}, B721:74--81, 2013.

\bibitem{Cullen:2013saa}
G.~Cullen, H.~van Deurzen, N.~Greiner, G.~Luisoni, P.~Mastrolia, et~al.
\newblock {Next-to-Leading-Order QCD Corrections to Higgs Boson Production Plus
  Three Jets in Gluon Fusion}.
\newblock {\em Phys.Rev.Lett.}, 111(13):131801, 2013.

\bibitem{Dixon:1996wi}
Lance~J. Dixon.
\newblock {Calculating scattering amplitudes efficiently}.
\newblock 1996.

\bibitem{Peskin:2011in}
Michael~E. Peskin.
\newblock {Simplifying Multi-Jet QCD Computation}.
\newblock 2011.

\bibitem{Buchmuller:1985jz}
W.~Buchmuller and D.~Wyler.
\newblock {Effective Lagrangian Analysis of New Interactions and Flavor
  Conservation}.
\newblock {\em Nucl.Phys.}, B268:621--653, 1986.

\bibitem{Dawson:1993qf}
S.~Dawson and R.~Kauffman.
\newblock {QCD corrections to Higgs boson production: nonleading terms in the
  heavy quark limit}.
\newblock {\em Phys.Rev.}, D49:2298--2309, 1994.

\bibitem{Gracey:2002he}
J.A. Gracey.
\newblock {Classification and one loop renormalization of dimension-six and
  dimension-eight operators in quantum gluodynamics}.
\newblock {\em Nucl.Phys.}, B634:192--208, 2002.

\bibitem{Melnikov:2000qh}
Kirill Melnikov and Timo~van Ritbergen.
\newblock {The Three loop relation between the MS-bar and the pole quark
  masses}.
\newblock {\em Phys.Lett.}, B482:99--108, 2000.

\bibitem{Pasechnik:2006du}
R.S. Pasechnik, O.V. Teryaev, and A.~Szczurek.
\newblock {Scalar Higgs boson production in a fusion of two off-shell gluons}.
\newblock {\em Eur.Phys.J.}, C47:429--435, 2006.

\bibitem{Collins:1978wz}
John~C. Collins, Frank Wilczek, and A.~Zee.
\newblock {Low-Energy Manifestations of Heavy Particles: Application to the
  Neutral Current}.
\newblock {\em Phys.Rev.}, D18:242, 1978.

\bibitem{Kilgore:2013gba}
William~B. Kilgore.
\newblock {One-Loop Single-Real-Emission Contributions to $pp\to H + X$ at
  Next-to-Next-to-Next-to-Leading Order}.
\newblock {\em Phys.Rev.}, D89:073008, 2014.

\bibitem{Gehrmann:2011aa}
T.~Gehrmann, M.~Jaquier, E.W.N. Glover, and A.~Koukoutsakis.
\newblock {Two-Loop QCD Corrections to the Helicity Amplitudes for $H \to$ 3
  partons}.
\newblock {\em JHEP}, 1202:056, 2012.

\bibitem{KlubergStern:1974rs}
H.~Kluberg-Stern and J.B. Zuber.
\newblock {Ward Identities and Some Clues to the Renormalization of Gauge
  Invariant Operators}.
\newblock {\em Phys.Rev.}, D12:467--481, 1975.

\bibitem{Tarrach:1981bi}
R.~Tarrach.
\newblock {The Renormalization of Ff}.
\newblock {\em Nucl.Phys.}, B196:45, 1982.

\bibitem{Grinstein:1988wz}
Benjamin Grinstein and Lisa Randall.
\newblock {The Renormalization of $g^{2}$}.
\newblock {\em Phys.Lett.}, B217:335, 1989.

\bibitem{Abbott:1980hw}
L.F. Abbott.
\newblock {The Background Field Method Beyond One Loop}.
\newblock {\em Nucl.Phys.}, B185:189, 1981.

\bibitem{Alloul:2013bka}
Adam Alloul, Neil~D. Christensen, CŽline Degrande, Claude Duhr, and Benjamin
  Fuks.
\newblock {FeynRules 2.0 - A complete toolbox for tree-level phenomenology}.
\newblock {\em Comput.Phys.Commun.}, 185:2250--2300, 2014.

\bibitem{Hahn:2000kx}
Thomas Hahn.
\newblock {Generating Feynman diagrams and amplitudes with FeynArts 3}.
\newblock {\em Comput.Phys.Commun.}, 140:418--431, 2001.

\bibitem{Hahn:1998yk}
T.~Hahn and M.~Perez-Victoria.
\newblock {Automatized one loop calculations in four-dimensions and
  D-dimensions}.
\newblock {\em Comput.Phys.Commun.}, 118:153--165, 1999.

\bibitem{Mertig:1990an}
R.~Mertig, M.~Bohm, and Ansgar Denner.
\newblock {FEYN CALC: Computer algebraic calculation of Feynman amplitudes}.
\newblock {\em Comput.Phys.Commun.}, 64:345--359, 1991.

\bibitem{Ellis:2007qk}
R.~Keith Ellis and Giulia Zanderighi.
\newblock {Scalar one-loop integrals for QCD}.
\newblock {\em JHEP}, 0802:002, 2008.

\bibitem{Schmidt:1997wr}
Carl~R. Schmidt.
\newblock {$H \to g g g (g q {\overline{q}}$) at two loops in the large $M_t$
  limit}.
\newblock {\em Phys.Lett.}, B413:391--395, 1997.

\bibitem{Harris:2001sx}
B.W. Harris and J.F. Owens.
\newblock {The Two cutoff phase space slicing method}.
\newblock {\em Phys.Rev.}, D65:094032, 2002.

\bibitem{Beenakker:1988bq}
W.~Beenakker, H.~Kuijf, W.L. van Neerven, and J.~Smith.
\newblock {QCD Corrections to Heavy Quark Production in p anti-p Collisions}.
\newblock {\em Phys.Rev.}, D40:54--82, 1989.

\bibitem{Dawson:1991au}
S.~Dawson and R.P. Kauffman.
\newblock {Higgs boson plus multi - jet rates at the SSC}.
\newblock {\em Phys.Rev.Lett.}, 68:2273--2276, 1992.

\bibitem{Kauffman:1996ix}
Russel~P. Kauffman, Satish~V. Desai, and Dipesh Risal.
\newblock {Production of a Higgs boson plus two jets in hadronic collisions}.
\newblock {\em Phys.Rev.}, D55:4005--4015, 1997.

\bibitem{Dixon:1993xd}
Lance~J. Dixon and Yael Shadmi.
\newblock {Testing gluon self-interactions in three jet events at hadron
  colliders}.
\newblock {\em Nucl.Phys.}, B423:3--32, 1994.

\bibitem{Dixon:2004za}
Lance~J. Dixon, E.W.~Nigel Glover, and Valentin~V. Khoze.
\newblock {MHV rules for Higgs plus multi-gluon amplitudes}.
\newblock {\em JHEP}, 0412:015, 2004.

\bibitem{Neill:2009mz}
Duff Neill.
\newblock {Analytic Virtual Corrections for Higgs Transverse Momentum Spectrum
  at ${\cal{O}}(\alpha_s^2/M_t^3)$ via Unitarity Methods}.
\newblock 2009.

\bibitem{Broedel:2012rc}
Johannes Broedel and Lance~J. Dixon.
\newblock {Color-kinematics duality and double-copy construction for amplitudes
  from higher-dimension operators}.
\newblock {\em JHEP}, 1210:091, 2012.

\bibitem{Cachazo:2004kj}
Freddy Cachazo, Peter Svrcek, and Edward Witten.
\newblock {MHV vertices and tree amplitudes in gauge theory}.
\newblock {\em JHEP}, 0409:006, 2004.

\bibitem{Alwall:2014hca}
J.~Alwall, R.~Frederix, S.~Frixione, V.~Hirschi, F.~Maltoni, et~al.
\newblock {The automated computation of tree-level and next-to-leading order
  differential cross sections, and their matching to parton shower
  simulations}.
\newblock {\em JHEP}, 1407:079, 2014.

\bibitem{Degrande:2011ua}
Celine Degrande, Claude Duhr, Benjamin Fuks, David Grellscheid, Olivier
  Mattelaer, et~al.
\newblock {UFO - The Universal FeynRules Output}.
\newblock {\em Comput.Phys.Commun.}, 183:1201--1214, 2012.

\bibitem{Owens:2012bv}
J.F. Owens, A.~Accardi, and W.~Melnitchouk.
\newblock {Global parton distributions with nuclear and finite-$Q^2$
  corrections}.
\newblock {\em Phys.Rev.}, D87(9):094012, 2013.

\bibitem{deFlorian:2011xf}
Daniel de~Florian, Giancarlo Ferrera, Massimiliano Grazzini, and Damiano
  Tommasini.
\newblock {Transverse-momentum resummation: Higgs boson production at the
  Tevatron and the LHC}.
\newblock {\em JHEP}, 1111:064, 2011.

\end{thebibliography}

\appendix
\section{Virtual Contributions}
\label{sec:appa}
Defining $V_i$, along with the logarithms and dilogarithms,
 as complex numbers, the one-loop $qg$ virtual contributions proportional to $C_1$ are \cite{Schmidt:1997wr},
\begin{eqnarray}
V_1&=& -{1\over\epsilon^2}\biggl[\biggl({m_H^2\over -S_{gq}}\biggr)^\epsilon
+\biggl({m_H^2\over -S_{g{\bar q}}}\biggr)^\epsilon\biggr]+{13\over 6\epsilon}
\biggl(
{m_H^2\over -S_{q{\bar q}}}\biggr)^\epsilon
\nonumber \\ &&
-\log\biggl(
{S_{gq}\over m_H^2}\biggr)
\log\biggl(
{S_{q{\bar q}}\over m_H^2}\biggr)
-\log\biggl(
{S_{g{\bar q}}\over m_H^2}\biggr)
\log\biggl(
{S_{q{\bar q}}\over m_H^2}\biggr)
-2Li_2\biggl(1-{S_{q{\bar q}}\over m_H^2}\biggr)
\nonumber \\ &&
-Li_2\biggl(1-{S_{g{q}}\over m_H^2}\biggr)
-Li_2\biggl(1-{S_{g{\bar q}}\over m_H^2}\biggr)
+{40\over 9}+{\pi^2\over 3}-{S_{q{\bar q}}\over 2 S_{g{\bar q}}}
\nonumber \\
V_2&=& \biggl[{1\over\epsilon^2}
+{3\over 2\epsilon}\biggr]\biggl({m_H^2\over- S_{q{\bar q}}}\biggr)^\epsilon
+\log\biggl(
{S_{gq}\over m_H^2}\biggr)
\log\biggl(
{S_{g{\bar q}}\over m_H^2}\biggr)
+Li_2\biggl(1-{S_{gq}\over m_H^2}\biggr)
\nonumber \\ &&
+Li_2\biggl(1-{S_{g{\bar q}}\over m_H^2}\biggr)
+4-{\pi^2\over 6}-{S_{q{\bar q}}\over S_{g{\bar q}}}
\nonumber \\
V_3&=& -{2\over 3\epsilon} \biggl({m_H^2\over -S_{q{\bar q}}}\biggr)^\epsilon -{10\over 9}\,.
\end{eqnarray}
These results are in agreement with Ref. \cite{Schmidt:1997wr}.   The results must be analytically
continued for timelike momentum invariants: $\log(S_{ij})\rightarrow \log(\mid S_{ij}\mid)+i\pi \theta(-S_{ij})$ and  $(-1)^\epsilon\rightarrow
1+i\pi\epsilon -{{\epsilon^2}\pi^2\over 2}$.

The one-loop $qg$ virtual contributions proportional to $C_5$ are (with $W_i$ complex),
\begin{eqnarray}
W_{1} & =&\frac{1}{\epsilon^{2}}\biggl[\
\biggl({m_H^2\over -S_{g{\bar q}}}\biggr)^\epsilon+
\biggl({m_H^2\over -S_{gq}}\biggr)^\epsilon\biggl]
+\frac{1}{\epsilon}\left[\frac{17}{6}\right] 
\nonumber \\
& & 
-\log\left(\frac{S_{gq}}{m_{H}^{2}}\right)
 -\frac{33}{18}\log\left(\frac{S_{g\bar{q}}}{m_{H}^{2}}\right)+\frac{121}{18}+\frac{1}{6}\frac{S_{g\bar{q}}}{S_{gq}}+\frac{1}{3}\frac{S_{q\bar{q}}}{S_{gq}}\nonumber \\
W_{2} & = &
-\frac{1}{\epsilon^{2}}\biggl({m_H^2\over -S_{q\bar q}}\biggr)^\epsilon
+\frac{1}{\epsilon}\left[-\frac{17}{6}\right]
+\log\left(\frac{S_{gq}}{m_{H}^{2}}\right)
 \quad+\frac{1}{3}\log\left(\frac{S_{g\bar{q}}}{m_{H}^{2}}\right)
\nonumber \\ &&
+\frac{3}{2}\log\left(\frac{S_{q\bar{q}}}{m_{H}^{2}}\right)
 -\frac{103}{18}
 +\frac{1}{3}\frac{S_{g{\bar q}}}{S_{g{q}}}+\frac{1}{6}\frac{S_{q\bar{q}}}{S_{gq}}\\
W_3&=& {2\over 3}\biggl[ {1\over\epsilon}-\log\biggl({-S_{q\bar{q}}\over m_H^2}\biggr)\biggr]
+{10\over 9}\quad .
\end{eqnarray}
This result is in disagreement with that of Ref. \cite{Neill:2009mz}.

The one-loop $gg$ contribution proportional to $C_1$ is,
\begin{eqnarray}
U_{1} & =&
-\frac{1}{\epsilon^{2}}\biggl[
\biggl({m_H^2\over -S_{12}}\biggr)^\epsilon
+
\biggl({m_H^2\over -S_{23}}\biggr)^\epsilon
+\biggl({m_H^2\over -S_{31}}\biggr)^\epsilon\biggr]\nonumber \\
  && \quad-\log\left(\frac{S_{23}}{m_{H}^{2}}\right)\log\left(\frac{S_{31}}{m_{H}^{2}}\right)
 -\log\left(\frac{S_{31}}{m_{H}^{2}}\right)\log\left(\frac{S_{12}}{m_{H}^{2}}\right)
 \nonumber \\
& & \quad-\log\left(\frac{S_{12}}{m_{H}^{2}}\right)\log\left(\frac{S_{23}}{m_{H}^{2}}\right)-2{\rm Li}_{2}\left(1-\frac{S_{12}}{m_{H}^{2}}\right)\nonumber \\
& & \quad-2{\rm Li}_{2}\left(1-\frac{S_{23}}{m_{H}^{2}}\right)-2{\rm Li}_{2}\left(1-\frac{S_{31}}{m_{H}^{2}}\right)
\, ,
\nonumber 
\end{eqnarray}
which agrees  with Eq. (11) of Ref. \cite{Schmidt:1997wr}.

The one-loop $gg$ contribution proportional to $C_3$ is,
\begin{eqnarray}
U_{3} & =&
 -\frac{3}{\epsilon^{2}
 \left(1-2\epsilon\right)}
 \left[\left(\frac{m_{H}^{2}}{-S_{12}}\right)^{\epsilon}
 +\left(\frac{m_{H}^{2}}{-S_{23}}\right)^{\epsilon}+\left(\frac{m_{H}^{2}}{-S_{31}}\right)^{\epsilon}\right]+O(\epsilon).
\end{eqnarray}

\section{NLO Real Emission - Quark Amplitudes}
\label{sec:appb}
\subsection{$q\bar q ggh$ amplitudes}
The contribution from $O_3$, to be multiplied by $C_3$, is
\begin{align}
im^{O_3} \left( q_-(1), g_-(2), g_-(3), \bar q_+ (4), h \right) &= -3 i g_s \frac{ \langle 12 \rangle \langle 23 \rangle \langle 31 \rangle} {\langle 14 \rangle}, \\
\label{O3qggqh}
im^{O_3} \left( q_-(1), g_-(2), g_+(3), \bar q_+ (4), h \right) &=0, \\
im^{O_3} \left( q_-(1), g_+(2), g_-(3), \bar q_+ (4), h \right) &=0, \\
\end{align}
Just like the $ggggh$ amplitudes in Section \ref{nloreal}, Eq. \eqref{O3qggqh} demonstrates non-interference with the $O_1$ amplitude in the soft Higgs limit. The $O_4$ operator contains two pairs of quark bilinears, so does not contribute to the $q\bar q ggh$ tree amplitude. The $O_5$ operator is easily shown to satisfy the operator relation
\begin{equation}
O_5 = O_4 + \partial^\alpha h G^{A}_{\alpha \nu} D^\beta G^{A\, \nu}_\beta ,
\end{equation}
up to total derivatives, which leads to the following contributions proportional to $p_H$, to be multiplied by $C_5$,
\begin{align}
&im^{O_5} \left( q_-(1), g_+(2), g_-(3), \bar q_+ (4), h \right) \nonumber \\
&= g_s^2 \left[
\frac{ i \langle 13 \rangle \langle 3 \slashed p_H 2] \langle 1 \slashed p_H 4] } {2 \langle 12 \rangle S_{23} }
- \frac { i [24] \langle 1 \slashed p_H 2] \langle 1 \slashed p_H 4] } {2 \langle 12 \rangle [23] [34] }
+ \frac {i [24] \langle 13 \rangle^2 } {\langle 12 \rangle S_{23} } p_H \cdot (p_2 + p_3 )
\right], \\
&im^{O_5} \left( q_-(1), g_-(2), g_+(3), \bar q_+ (4), h \right) \nonumber \\
&= g_s^2 \left[ \frac { i \langle 12 \rangle [34]} {S_{23} [12] \langle 34 \rangle } \left( [13] \langle 34 \rangle p_H \cdot p_3 - [12] \langle 24 \rangle p_H \cdot p_2 \right) \right. \nonumber \\
&\left. - \frac i 2 \frac { \langle 2 \slashed p_H 3] } { \langle 34 \rangle [12] S_{23} } \left(S_{13} S_{34} - S_{24} S_{12} + S_{23} S_{34} - S_{23} S_{12} \right)
\right], \\
&im^{O_5} \left( q_-(1), g_-(2), g_-(3), \bar q_+ (4), h \right) \nonumber \\
&= g_s^2 \left[ -\frac i 2 \frac{ ( S_{12} + S_{13} + S_{23} ) \langle 3 \slashed p_H 4] } {2 [12][23]} - \frac {i \langle 1 \slashed p_H 4] \langle 2 \slashed p_H 4]} {2 [23][34]}
\right]
\end{align}
\subsection{$q\bar q q\bar q$ and $q\bar q Q\bar Q$ amplitudes}
The $O_3$ amplitude vanishes at tree-level due to the absence of the $ggh$ vertex.
For $O_4$ and $O_5$, we define
\begin{align}
f_4 (p_1, p_2, p_3, p_4) &= 2i \langle 14 \rangle [32], \\
f_5 (p_1, p_2, p_3, p_4) &= \frac i 2 \left( \frac 1 {S_{12}} + \frac 1 {S_{34}} \right) \nonumber \\
& \quad \left[ \langle 1 \slashed p_H 2] \langle 4 \slashed p_H 3] + \langle 14 \rangle [23] (p_1+p_2) \cdot (p_3+p_4) \right].
\end{align}
The amplitudes for $O_i$, $i=4,5$, are
\begin{align}
im^{O_i} \left( q_-^{c_1}(1), \bar q_+^{c_2}(2), Q_+^{c_3}(3), \bar Q_-^{c_4}, h \right) &= 
im^{O_i} \left( q_-^{c_1}(1), \bar q_+^{c_2}(2), q_+^{c_3}(3), \bar q_-^{c_4}, h \right) \nonumber \\
&= g_s^2 f_i (p_1, p_2, p_3, p_4) \sum_A T^A_{c_1 c_2} T^A_{c_3 c_4},  \\
im^{O_i} \left( q_-^{c_1}(1), \bar q_+^{c_2}(2), q_-^{c_3}(3), \bar q_+^{c_4}, h \right) &= g_s^2 f_i (p_1, p_2, p_4, p_3) \sum_A T^A_{c_1 c_2} T^A_{c_3 c_4}, \nonumber \\
& \quad + f_i (p_3, p_2, p_4, p_1 ) \sum_A T^A_{c_3 c_2} T^A_{c_1 c_4}\, ,
\end{align}
where $q$ and $Q$ represent different flavor quarks.

\end{document}